\documentclass[12pt,preprint]{aastex}

\shorttitle{Asphericity of the Sun}
\shortauthors{Basu et al.}

\newcommand\ea{et al.}

\begin{document}

\title{Structure of the near-surface layers of the Sun: asphericity and
time variation}

\author{Sarbani Basu}
\affil{Astronomy Department, Yale University, P. O. Box 208101,
New Haven CT 06520-8101, U. S. A.}
\email{sarbani.basu@yale.edu}
\author{H. M. Antia}
\affil{Tata Institute of Fundamental Research,
Homi Bhabha Road, Mumbai 400005, India}
\email{antia@tifr.res.in}
\and
\author{Richard S. Bogart}
\affil{Stanford University, CSSA - HEPL A202, Stanford, CA 94305-4085, U. S. A.}
\email{rbogart@spd.aas.org}

\begin{abstract}
We present results on the structure of the near-surface layers
of the Sun obtained by inverting frequencies
of high-degree solar modes from ``ring diagrams''.
We have results for eight epochs between June 1996 and October 2003.
The frequencies for each epoch were obtained from ring diagrams constructed
from MDI Dopplergrams spanning complete Carrington rotations.
We find that there is a substantial latitudinal variation of
both sound speed and the adiabatic index $\Gamma_1$ in the outer 2\% of
the Sun. We find that both the sound-speed and
$\Gamma_1$ profiles change with changes in the  level of solar activity.
In addition, we also study differences between the northern and southern hemispheres
of the Sun and find a small asymmetry that appears to reflect the
difference in magnetic activity between the two hemispheres.
\end{abstract}


\section{Introduction}
\label{sec:intro}

Seismic data have been successfully used to determine the
solar interior structure (e.g., Gough {\it et al.}~1996).
Departures from spherical symmetry  have been less
well studied. Among the more  detailed studies is one by Antia  {\it et al.}~(2001).
This study showed that the solar sound-speed profile is not
spherically symmetric, and that it depends on both radius and latitude.
Helioseismic determinations of solar structure --- both the spherically
symmetric and the latitudinally dependent parts ---  are  restricted to
the deeper layers of
the Sun ($ r < 0.95 R_\odot$) because of the limitations of the
mode sets that are
routinely determined by the Michelson Doppler Imager (MDI)
on board the Solar and Heliospheric
Observatory (SOHO) and from the ground-based
Global Oscillations Network Group (GONG). 
These data sets provide frequencies for {\it p}-modes with degree
$\ell \le 190$ and {\it f}-modes with degree  $\ell\le 300$ for the most part,
and these low- and intermediate-degree
modes are not very useful in determining the structure of the
near-surface layers of  the Sun. However, 
the {\it f}-modes, which are confined to the outermost layers of the Sun,
are more sensitive to rotation than structure,  and these provide
reliable determinations of solar rotation to fairly shallow depths.
Solar rotation and other large-scale flows are known to have a 
strong latitudinal dependence (e.g., Thompson {\it et al.}~1996; Schou {\it et al.}~1998);
this dependence is also known to show a time variation that 
is correlated with solar activity (e.g.,  Schou 1999;
Basu \& Antia 2000, 2002, 2003;
Howe {\it et al.}~2000, 2004a). Changes in the latitudinal variation
of structure in near-surface regions are, however, difficult to study because of the lack of
availability of frequencies for high-degree modes. 

The changes in frequencies of intermediate-degree global modes
that are routinely determined from MDI and GONG data
show that there are no clearly observable changes in
solar structure below $0.95R_\odot$ (Eff-Darwich {\it et al.}~2002; Basu 2002).
There is tentative and indirect evidence that there are
indeed solar-cycle related changes in the adiabatic index $\Gamma_1$ in the
shallower layers of the Sun, around the He II ionization zone,
i.e. $\approx 0.98R_\odot$,
(Basu \& Mandel 2004; Verner {\it et al.}~2006). Chou \& Serebryanskiy (2005)
find some marginal evidence for temporal variations near the base
of the convection zone,
which contradicts earlier negative results (Eff-Darwich {\it et al.}~2002;
Basu 2002).
Similarly, inversions of the
frequency splittings of these modes do not show any significant time variations
in the latitudinal distribution of sound speed and density below
$0.95R_\odot$ (Antia {\it et al.}~2001, 2003). However, the ``surface term''
from the inversion
(see \S~\ref{sec:inv}) changes with time,
implying that there could be changes in the structure
closer to the solar surface.  The surface term also shows a distinct
correlation with the distribution of magnetic fields at the solar surface
(Fig.~8 of Antia {\it et al.}~2001). This leads us to
believe that we may be able to detect changes in the structure of the
outer layers of the Sun associated with magnetic
activity changes by using  frequencies of high-degree modes.
It is already known that frequencies of high-degree
modes vary with local magnetic field strength (e.g., Hindman {\it et al.}~2000;
Rajaguru {\it et al.}~2001; Howe {\it et al.}~2004b),
and in principle, these frequency differences can be inverted to infer
corresponding variations in solar structure or other properties.
Earlier studies have shown that the profiles of sound speed and $\Gamma_1$
in regions of high magnetic activity differ from those
in quiet regions (Kosovichev {\it et al.}~2000, 2001; Basu, Antia \& Bogart~2004).
These studies show that for magnetically active regions, sound speed and
adiabatic index $\Gamma_1$ are lower than that of quiet regions
in the immediate sub-surface layers (about the outer seven Mm or so). 

If there are indeed changes in the outer layers of the Sun,
we need frequencies
of high-degree modes to detect these changes. Unfortunately,
it is very difficult to determine the frequencies of very high degree
modes through global analysis
(e.g., Korzennik {\it et al.}~2004 and references therein) and as a result
neither GONG nor MDI provide data on high degree modes as standard 
products, and frequency sets with high-degree modes are not available
for different epochs.
High-degree solar  modes ($l\ge150$)
which are trapped in the outer parts of the solar envelope have lifetimes
that are much shorter than the sound travel time around the Sun and hence the
characteristics of these modes are mainly determined by the average
conditions in the local neighborhood rather than those of the
entire sphere. The properties of these modes are
better determined by local helioseismic techniques.
Ring-diagram (plane-wave $k$--$\nu$) and time-distance analysis (Duvall
\ea~1993)
are examples of such local helioseismic techniques.
Frequencies obtained
from ring-diagram analysis have been successfully used to determine the
difference in structure between active and quiet regions of the
Sun (Basu, Antia \& Bogart 2004).

Since high-degree global-mode data
are not standard products of the GONG or MDI projects, we use a ring-diagram analysis
to obtain solar oscillation frequencies
of different latitudinal zones of the Sun at different times. This technique allows
us to do a differential analysis of the structure between different regions.
We use these frequencies to determine the differences in structure
between the equator and higher latitudes. 
Since we are interested in the near-surface layers where ionization occurs,
we need to invert for both the 
sound speed ($c$) and the adiabatic index ($\Gamma_1$)
in order to be able to interpret the results, since any temperature change
in the ionization zone will also change $\Gamma_1$ there.
To study
how the latitudinal dependence of structure changes with time and activity,
we analyze
several epochs corresponding to different solar activity levels. 
The differences in structure between the equator and higher
latitudes are expected to be less than those found between strong
active regions and quiet regions, because of the smaller differences
in the zonally averaged magnetic fields, nevertheless
we expect the differences to be detectable.

The rest of the paper is organized as follows: In \S~\ref{sec:data}
we discuss the analysis technique and the regions studied;
we give a brief description of the inversion process in \S~\ref{sec:inv};
we describe and discuss our results in \S~\ref{sec:res}, and state our
conclusions in \S~\ref{sec:conclu}.

\section{Data Analysis}
\label{sec:data}

The basic data set consists of the
full-disk Dopplergrams obtained by  the MDI instrument.
Ring diagrams (Hill 1988) are three-dimensional (3D) power spectra of short-wavelength
modes in a small region of the Sun. High-degree (short-wavelength) modes
can be approximated as plane waves over a small area of the Sun as long as
the horizontal wavelength of the modes is much smaller than the solar
radius.  Ring diagrams are obtained from a time series of Dopplergrams of
a specific area of the Sun that are usually tracked with the mean rotation velocity. The 3D
Fourier transform of this time series gives the power spectra. These power
spectra are  referred to as ring diagrams because of the
characteristic ring-like shape of regions where the power is concentrated
in cuts of constant
temporal frequency, reflecting the near azimuthal symmetry of the
power in $\bf k$ space. A detailed description of the ring-diagram technique
is given by Patron \ea\ (1997) and Basu \ea\ (1999).
Ring-diagram analysis
has the advantage over global-mode analysis that it can be used to study
both the non-axisymmetric component of the
structure and dynamics of the Sun, as well as the anti-symmetric component 
of these quantities between the solar northern and southern hemispheres.

For this work we have chosen to analyze MDI Doppler data from eight complete Carrington
rotations. The time intervals chosen 
were dictated by the times at which MDI was in its
``Dynamics Program'' observing mode, for which full-disc Dopplergrams at
a one-minute cadence are available nearly continuously (at duty cycles
of at least 85\%) over periods as long as the analysis interval,
in this case at least one full Carrington rotation. 
The data intervals selected are listed in Table~1, which also gives the
average radio flux at 10.7 cm, a measure of the level
of solar activity, during each analysis period.

The data used  were not tracked at the photospheric rotation rate as in usual
ring-diagram
analysis. Instead, they were untracked in the sense that the longitudes were
referred to the central meridian of each observation. Corrections were
made, however, for the drift of the
spacecraft in heliographic latitude over the course of the analysis periods.
The analysis interval in each case
was 39,936 minutes ($24\times1664$ min), centered on the time of central
meridian crossing of Carrington longitude $180^\circ$ of the appropriate rotation as
viewed from SOHO. (For CR 2009, the analysis interval was centered on the central meridian
crossing of longitude $240^\circ$, about 4.5 days before the middle of the
rotation, in order to take better advantage of the most complete data
coverage from MDI.)  For each Carrington rotation we analyzed thirteen
latitude bands, each of width $15^\circ$, and with a spacing of
$7.5^\circ$, from $45^\circ$S to $45^\circ$N.
By comparing the results with those obtained using tracked data averaged
over the entire Carrington rotation, we have verified that the results
are not significantly affected. The untracked regions give us higher
frequency resolution because of their longer duration.

Since magnetic activity is a largely local phenomenon, we
have calculated the value of the ``Magnetic Activity Index'' (MAI)
of each set to give us a measure of the magnetic activity in each latitude zone.
The 10.7 cm radio flux on the other hand is related to total activity
averaged over the
entire visible hemisphere. The MAI is calculated by integrating the
unsigned magnetic field values
within the same regions and over the same intervals used to calculate the power
spectra, using available 96-minute MDI magnetograms. This index is a measure 
only of the strong field (fields less than 50 Gauss are set to zero to avoid contamination 
by  zero-level errors and residual noise).
The same temporal and spatial apodizations were
used.  Details of how the MAI is calculated are given by
Basu {\it et al.}~(2004). The MAI's show the
usual butterfly diagram pattern, and their values range from a minimum of 0.3~G to 
a maximum of about 24~G.

The highly elliptical untracked ring spectra were fitted using the same
13-parameter fit we have used for tracked ones (Basu \& Antia, 1999),
with suitable adjustments for the very large values of the $U_x$
parameter reflecting the advection of the waves by solar rotation
in addition to local proper motion.

\section{Inversion Techniques}
\label{sec:inv}

Inversion for solar structure is  complicated because
the problem
is inherently non-linear. The inversion
generally  proceeds through a linearization
of the equations of stellar oscillations,
using their variational formulation, around a known reference
model (see e.g., Dziembowski {\it et al.}~1990;
D\"appen  {\it et al.}~1991; Antia \& Basu 1994;
Dziembowski {\it et al.}~1994, etc.).  The differences  between the
structure of the Sun and the reference model are then related to the
differences in the frequencies of the Sun and the model by kernels.
Nonadiabatic effects and other errors in modeling the surface layers
give rise to frequency shifts (Cox \& Kidman
1984; Balmforth 1992) which are not accounted for by the variational
principle.
In the absence of any reliable formulation, these
effects have been taken into account in an {\it ad hoc} manner by
including an arbitrary function of frequency in the variational
formulation (Dziembowski {\it et al.}~1990).

The fractional change in frequency of a mode can be expressed
in terms of fractional
changes in the structure of model characteristics,
for example, the adiabatic sound speed $c$ and density $\rho$,
and a surface term.
The frequency differences can be written in the form:
\begin{equation}
{\delta \nu_i \over \nu_i}
=  \int_0^{R_\odot} K_{c^2,\rho}^i(r){ \delta c^2(r) \over c^2(r)}\; {\rm d} r +
 \int_0^{R_\odot} K_{\rho,c^2}^i(r) {\delta \rho(r)\over \rho(r)}\; {\rm d} r
 +{F_{\rm surf}(\nu_i)\over I_i}
\label{eq:inv}
\end{equation}
(e.g., Dziembowski {\it et al.}~1990).  Here $\delta \nu_i$ is the
difference in the frequency $\nu_i$ of the $i$th mode between the
data and the reference model, where $i$ represents the pair ($n,l$), $n$ being the
radial order and $l$ the degree.
The kernels $K_{c^2, \rho}^i$
and $K_{\rho, c^2}^i$ are known functions that
relate the changes in frequency to the changes in the squared sound speed $c^2$ and 
density $\rho$
respectively, and $I_i$ is the mode inertia.
Instead of $(c^2, \rho)$, other pairs of functions may be used, such
as density and adiabatic index $\Gamma_1$.
The term $F_{\rm surf}$ is the ``surface term'',  and takes into account
the near-surface errors in modeling the structure. This term  also contains contributions
from the very shallow  layers of the Sun that cannot be probed by the
mode set used.

In this work we determine the relative differences in the squared sound speed,
$\delta c^2/c^2$, and
the adiabatic index  $\delta\Gamma_1/\Gamma_1$, as functions of depth,
between the solar equator and higher latitudes.
In order to minimize systematic errors in the structure inversions, we
invert the frequency differences
between different parts of the Sun, rather than frequency differences between
a region of the Sun and a solar model. To study the latitudinal structure
we study differences in solar structure between the higher latitudes and the equator.
The  main
reason for inverting the differences between two sets of solar frequencies is
that the structure of the Sun could differ by a  very large amount from solar models
in the near-surface regions. There
are usually many assumptions involved in obtaining the
near-surface structure of a solar model,
such as the use of the mixing-length formalism to
treat convection, the diffusion approximation to treat radiation, and
ignoring turbulent pressure; these
usually break down close to the surface.  Modeling errors may result in
large differences in structure between the models and the inverted data in the
near-surface layers, and these differences are not always in the linear
regime. As a result, Eq.~\ref{eq:inv} may not hold, and its use could result in
systematic errors. The differences between the structure at different
latitudes, on the other hand, are expected to be small, and we can thus
use Eq.~\ref{eq:inv} without introducing systematic errors in the results.
Another reason for doing a differential analysis is that the non-uniformities
in the MDI image geometry are not the same in all Dynamics campaigns.
The errors due to these changes are reduced when
contemporaneous frequencies are subtracted and used in Eq.~\ref{eq:inv}.
Also, the analysis technique itself involves certain approximations.
Since the spherical solar surface layers are modeled as being plane-parallel,
subtracting the mode frequencies removes some of the geometric errors
common to both sets of frequencies.  Of course we still need to use a
solar model to determine the kernels for the inversion.

Equation~(\ref{eq:inv}) constitutes the inverse problem
that must be solved to infer
the differences in structure between the solar equator and higher latitudes.
It can be inverted using a variety of techniques.
We carried out the inversions
using the Subtractive Optimally Localized Averages (SOLA) technique
(Pijpers \& Thompson 1992; 1994)
and the Regularized Least Squares (RLS) technique.
Details of how SOLA inversions are carried out and
how various parameters of the inversion are selected were given by
Rabello-Soares, Basu \& Christensen-Dalsgaard  (1999).
Details on RLS inversions and parameter
selections were provided by Antia \& Basu (1994)
and Basu \& Thompson (1996). Given the complementary nature of the
RLS and SOLA inversions (see Sekii 1997 for a discussion),
we can be more confident of the results if the two inversions
agree.

It must be noted that the kernels used in the inversions
were derived in the absence of magnetic field effects.
In regions with magnetic fields, there are two ways
frequencies can change. One is through the direct effect of magnetic fields
on the waves,
i.e., through the additional restoring force provided by the field;
the second is through the change in structure caused by the magnetic
field. The magnitude of these effects depends on the strength and
orientation of the magnetic field. To order of magnitude, the
relative change in frequency, or the effective squared wave speed,
can be expected to be of the order of  $v_A^2/c^2$,
where $v_A$ is the Alfv\'en speed.
Without any additional information it is not possible to distinguish
between the direct effect on frequencies of the magnetic field and the
indirect effect through modification of the structure. 

\section{Results}
\label{sec:res}

\subsection{Frequency differences}
\label{subsec:freqdif}

\begin{figure}
\plotone{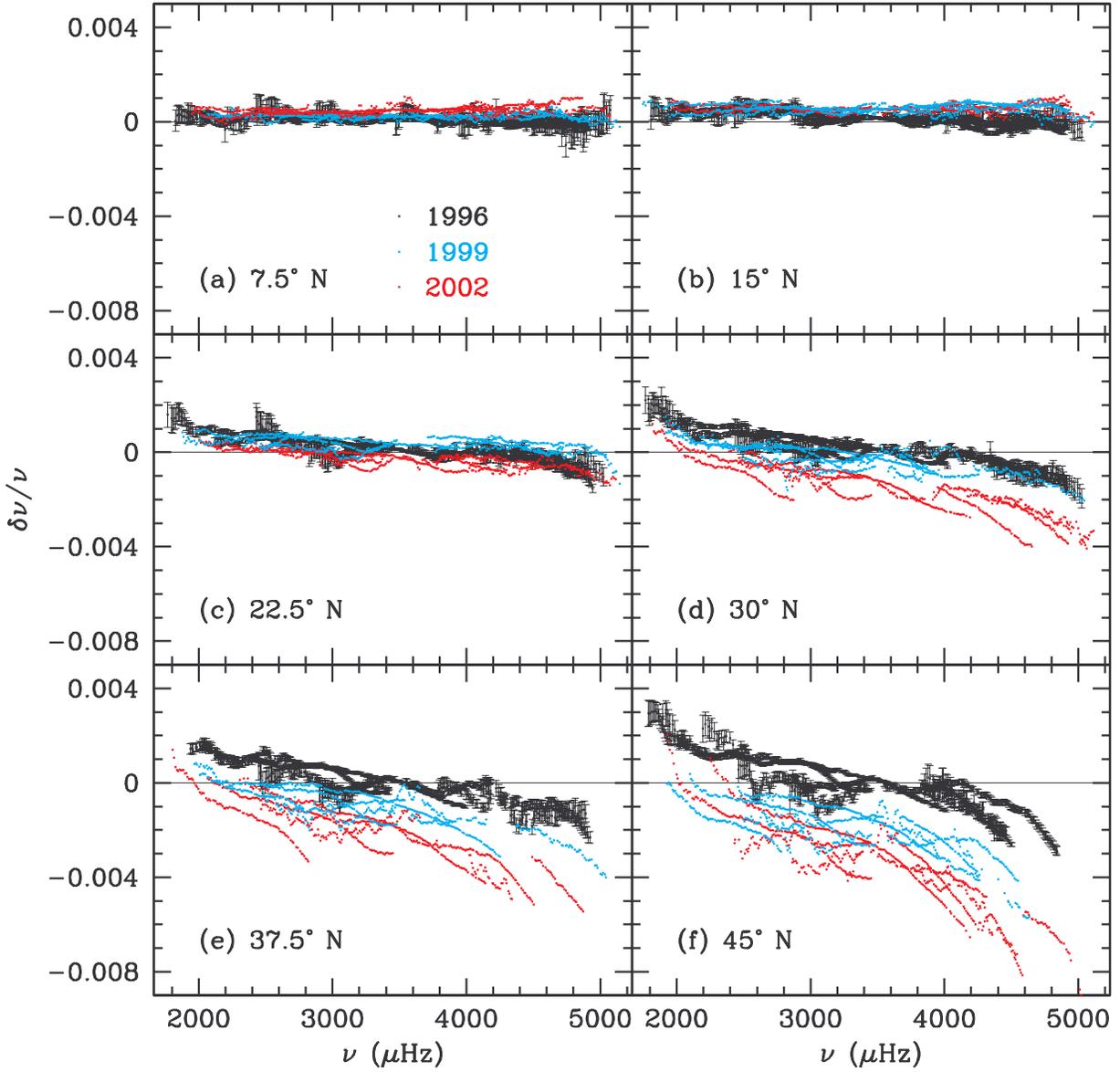}
\caption{The frequency differences between the equator and higher latitudes
in the northern hemisphere of the Sun for three different epochs. The epochs
are labeled by calendar year. The
differences are in the sense (Higher latitude $-$ Equator). For the
sake of clarity, the error bars are plotted for only one set; they are
similar for the other sets.
}
\label{fig:freqdif}
\end{figure}

Our results show clear frequency differences between the higher latitudes
and the equatorial region 
of the Sun. We also find that these differences change with time.
Fig.~\ref{fig:freqdif}
shows some of the frequency differences between different northern latitudes
and the equator for
three Carrington rotations in different years. The frequency differences
are significant in all cases. Similar results are seen for other years,
and also for latitude zones in the southern hemisphere. 

Since we find fairly substantial latitudinal effects, it is important to
examine possible systematic errors:  we expect systematic
errors in the frequencies obtained from high-latitude ring spectra due to
purely geometric projection effects. This effect
should also be present in frequency differences between
regions along the equator but at different longitudes with respect
to the disk center (neglecting smaller effects due to the non-zero
latitude of the observer), and we can use this fact to assess the
influence of projection effects on the mode frequencies.
If we assume that all the frequency differences between higher-latitude
regions and the equator are a result of projection effects, we should find
similar frequency differences between the disc center and
regions on the equator but away from the disc center, i.e., at different
central-meridian longitudes.
In order to test this, we determined the frequency
differences between the central meridian and other
longitudes at the solar equator, i.e. at different positions with
respect to the central meridian.  For this purpose
we analyzed two sets of quiet-Sun data, one from 1996 (CR 1910)
and the other from 1998 (CR 1932).
The data analysis was similar to that described in \S~\ref{sec:data}.
The power
spectra for all regions were constructed from 39,936 minutes of data,
and hence each power spectrum covered data for the same time interval
and the same integrated latitude band on the Sun, but at different
angles to the line of sight. In the absence of projection effects
we would expect the frequencies at zero longitude (the central meridian)
to be the same as those at other longitudes.  We did find systematic
differences between the higher longitude sets and the zero longitude set,
which show similar trend as that for the different latitude zones, with the largest
differences being for the maximum longitude separation of $45^\circ$;
they can be seen in Fig.~\ref{fig:freqcor}. But we also see that the
frequency differences in longitude are much smaller than those in
latitude
shown in Fig.~\ref{fig:freqdif}. Thus, we can be reasonably confident
that the observed frequency differences between different latitudes
and the equator are not simply due to projection effects.

\subsection{Radial Dependence of the Results}
\label{subsec:radial}

The frequency differences between the higher latitudes and the equator
 were inverted to determine the corresponding differences
in structure.
The inversion results are 
reliable over the radius range $0.975$--$0.996 R_\odot$, but  are
most reliable in terms of stability with respect to changes in
inversion parameters  between $0.985$ and $0.995 R_\odot$. A sample of the
results is shown in Fig.~\ref{fig:sample}. Note the agreement between
SOLA and RLS inversions.
We find that in most cases the two inversions give the same results
within the errors.
There are some differences for a few high-latitude cases,
but the agreement is generally good. 
In order to quantify the differences between the SOLA and RLS results, we
have determined the root-mean-squared (rms) differences, normalized
by the errors, between them. Figure~\ref{fig:histo}
shows the histograms of the differences for the radius ranges 
$0.975$--$0.996 R_\odot$ and $0.985$--$0.994 R_\odot$.
The distributions are consistent with random noise.

The inversion results show that for all the epochs studied, there are considerable
differences in sound speed  between the solar equator and the
higher latitudes. The averages of the northern and southern hemisphere
results are shown in Fig.~\ref{fig:csq}.
The solar sound speed varies as a function of latitude
at all epochs, though it appears that its latitudinal dependence
also varies at different epochs. The largest
differences are seen for the latitude of $45^\circ$, followed by
$37.5^\circ$. These latitudes generally have a higher sound speed than the
equator, particularly above about $0.985R_\odot$ or so. There are some signs that the
trend is altered closer to the surface. 

The $\Gamma_1$ differences between the equator and higher latitudes
displayed in Fig.~\ref{fig:gam} also
show a trend similar to the sound-speed differences, with larger values
of $\Gamma_1$ at higher latitudes than the equator.
In the case of $\Gamma_1$, these differences are more pronounced,
and we can also perceive a clear trend with time. 
The differences between
the equator and the higher latitudes appear to increase with
the level of solar activity. This is discussed further in \S~\ref{subsec:lat}.

If we concentrate on the structural changes that can be caused
by the presence of magnetic fields, we find that
it is relatively easy to change the sound speed: it can change if
the  gas pressure, temperature, or the mean molecular weight $\mu$ changes.
Gas pressure changes in response to the additional magnetic pressure,
the total pressure $P_{\rm tot}$ satisfying hydrostatic equilibrium
conditions being the
sum of the gas pressure and magnetic pressure, so that gas
pressure decreases when magnetic fields are present. 
Changing the adiabatic index $\Gamma_1$ is difficult, however.
It is determined
by the equation of state, and is expected to be close to 5/3 except in the
ionization zones, where it is lower.
Merely  changing pressure, density $\rho$ or temperature $T$ does not
change $\Gamma_1$ everywhere. Changes in $\Gamma_1$ can
be caused by shifts in the positions of the
ionization zones due to changes in the temperature profile.
But as pointed out above, magnetic effects are not manifested
only by changes in thermal structure. Frequency shifts also occur
because of the direct effect of magnetic fields in
changing the restoring forces. 
Part of the observed frequency shifts could be mistakenly
interpreted as changes in $\Gamma_1$  and $c^2$, since the inversion kernels
do not take the direct effect of the magnetic fields into account.

Estimating the change in temperature from the changes in $c^2$ and $\Gamma_1$ is
not straightforward. For an ideal gas, changes in $c^2$ and $\Gamma_1$ are
related by
\begin{equation}
{\delta c^2\over c^2}-{\delta \Gamma_1\over \Gamma_1}=\delta\ln {T\over \mu}.
\label{eq:ideal}
\end{equation} 
In the deeper layers of the Sun the gas is fully ionized, hence 
$\Gamma_1$ and $\mu$ are constant, and changes in $c^2$ can be related directly to changes in
$T$.
But the state in the ionization zones is more complicated. Ionization causes
both $\mu$ and $\Gamma_1$ to change, and hence it is not straightforward 
to  use the above equation to
relate changes in $c^2$ and $\Gamma_1$ to changes in temperature. 
The estimation of temperature perturbations is hampered by the fact that in the
ionization zones, changes
in $T$ will give rise to non-zero changes in $\mu$. We have no way to
directly determine changes in $\mu$. We could of course estimate how much
$\mu$ would change for a given change in $T$ using solar models or the
detailed equation of state.
But even though we may not be able to use the perturbations in $c^2$ and
$\Gamma_1$ to determine the temperature changes easily, we know that they
are always related by
\begin{equation}
{\delta c^2\over c^2}-{\delta \Gamma_1\over \Gamma_1} = \delta\ln {P\over \rho}.
\label{eq:c2diff}
\end{equation}
Thus the difference between the changes in $c^2$ and $\Gamma_1$
is still a significant quantity.

Figure~\ref{fig:dif} shows the differences between the $c^2$ and $\Gamma_1$
perturbation profiles.
We see that the $\Gamma_1$ differences do not account for all the sound-speed
difference, thus $P/\rho$ must change as well.  The errors in the results
are substantial. Nevertheless, we can see that at the high activity epoch,
the profile variations are significant at high latitudes.

\subsection{The surface term}
\label{subsec:sur}

Global-mode inversions for asphericity showed that the surface term of the
inversions follow the pattern of the surface magnetic field (Antia {\it et al.}~2001).
We have therefore, looked into how the surface term varies with MAI, our local 
magnetic field index. In Fig.~\ref{fig:mai} we show surface term contours
overplotted on $\delta {\rm MAI}$, the MAI difference between the different  regions
studied and the equator.  Note that the surface term is positive for positive MAI  
differences (i.e., when the magnetic field of the region is higher than that
of the equator), and negative when the MAI difference is negative. However,
the linear correlation coefficient of the surface term with MAI difference
is not very high, it is 0.38 when all data sets are used, but rises to
0.86 when only sets with $|\delta {\rm MAI} |> 5$ G are used.
The fact that there is still a good correlation implies that there 
are changes in layers shallower than those we can 
resolve in this investigation.  

The surface term also correlates with the global activity index. In particular, 
surface term from high-latitude inversions show a negative correlation
with the 10.7 cm flux, while those from low latitudes show a positive correlation.
The correlation coefficient at latitude of $30^\circ$ and higher is $-0.75$. The coefficient for
$15^\circ$ and lower is $0.45$, the lower absolute value of the
 correlation is probably due to the
fact that the magnetic field at the equator changes too as activity increases.  Since we are
doing a differential study with respect to the equator, and because  the equatorial and the
low-latitude surface terms behave the same way, the overall effect is reduced.

\subsection{Dependence of structure  on latitude, time, and activity}
\label{subsec:lat}

\subsubsection{Asphericity of structure}

In order to better determine the dependence of the results on latitude, we
concentrate on the averaged results in the radius ranges
$0.984$--$0.989 R_\odot$ and $0.989$--$0.994 R_\odot$.
Fig.~\ref{fig:lat} shows the averaged results for the two ranges
plotted as a function of latitude. We can see very clearly that
for the highest latitudes studied ($37.5^\circ$ and $45^\circ$)
both sound speed and $\Gamma_1$ are always higher at the 
higher latitudes than at the equator. 
At intermediate
latitudes they are generally lower than
at the equator. This behavior is seen most clearly in $\Gamma_1$.
At low latitudes (below $30^\circ$), $\Gamma_1$ is consistently lower than
that at the equator, while it is higher at latitudes
above $30^\circ$. At $30^\circ$, the results vary depending 
on the radius range and the epoch. 
The behavior of the sound speed differences is a bit more complicated, though
in the outer radius range sound speed and $\Gamma_1$ behave in a similar manner.
In all cases there
appear to be strong variations with time, with the differences increasing
during epochs of higher solar activity. The difference is particularly clear
for $\Gamma_1$. The $\Gamma_1$ differences become more negative below $30^\circ$
with time as solar activity increases, while the differences above $30^\circ$
become more positive. The decrease in $\Gamma_1$ in the outer layers at low
latitudes is what we would expect if the strength of magnetic fields
increased there; this is indeed what happens as the solar cycle progresses.

\begin{figure}
\plotone{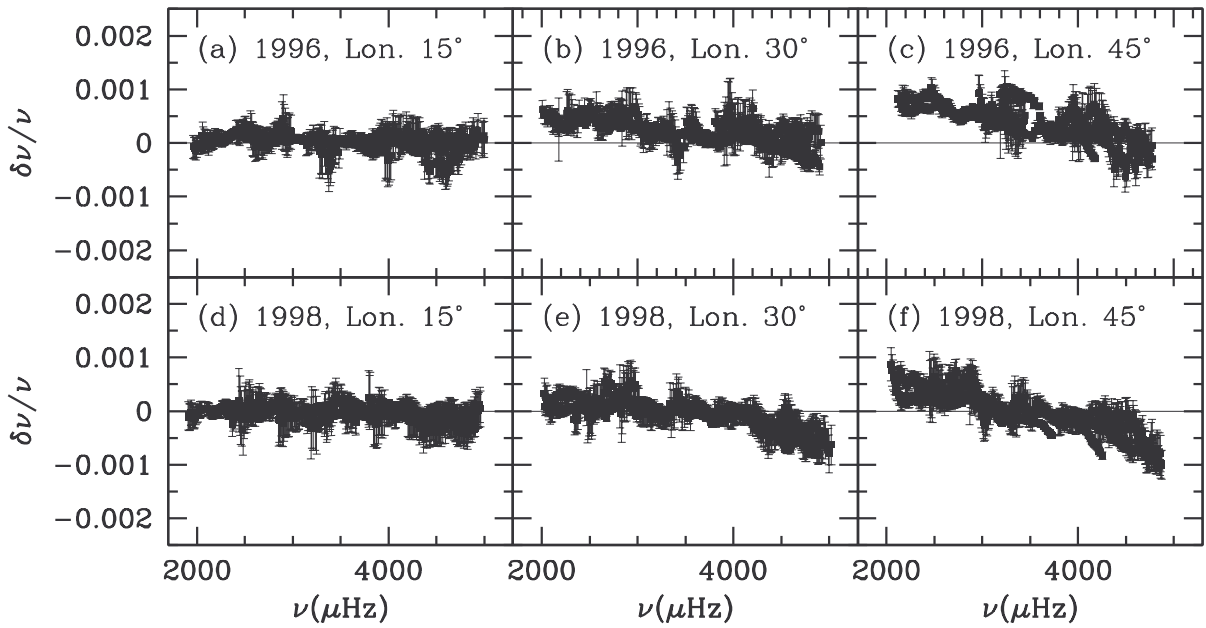}
\caption{The frequency differences between regions on and off central meridian
at the solar equator at two different times. The averages of the differences
for both east and west longitudes is shown for each separation.
Although there are some systematic effects, they are small
compared to the latitudinal differences shown in Fig.~\ref{fig:freqdif}
(note the scale difference).}
\label{fig:freqcor}
\end{figure}

\begin{figure}
\plotone{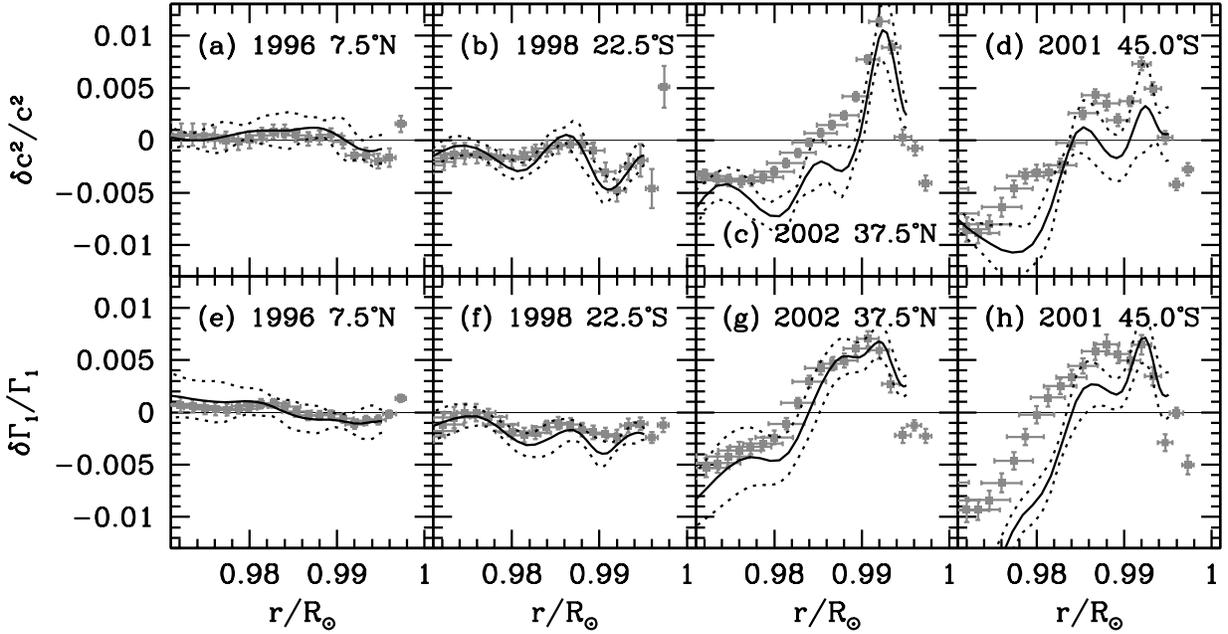}
\caption{Sample of the inversion results for differences in sound speed and
adiabatic index $\Gamma_1$ for selected data sets. Each panel is
labeled by the year of the data set and the latitude being compared with
the equator. The gray points are SOLA results; the vertical error bars 
correspond to $1\sigma$ errors, and the horizontal error bars are a
measure of the resolution of the inversions. The continuous lines
are RLS results with the dotted lines showing the $1\sigma$ errors.
All the differences are in the sense (Higher latitude $-$ Equator).
}
\label{fig:sample}
\end{figure}

\begin{figure}
\plotone{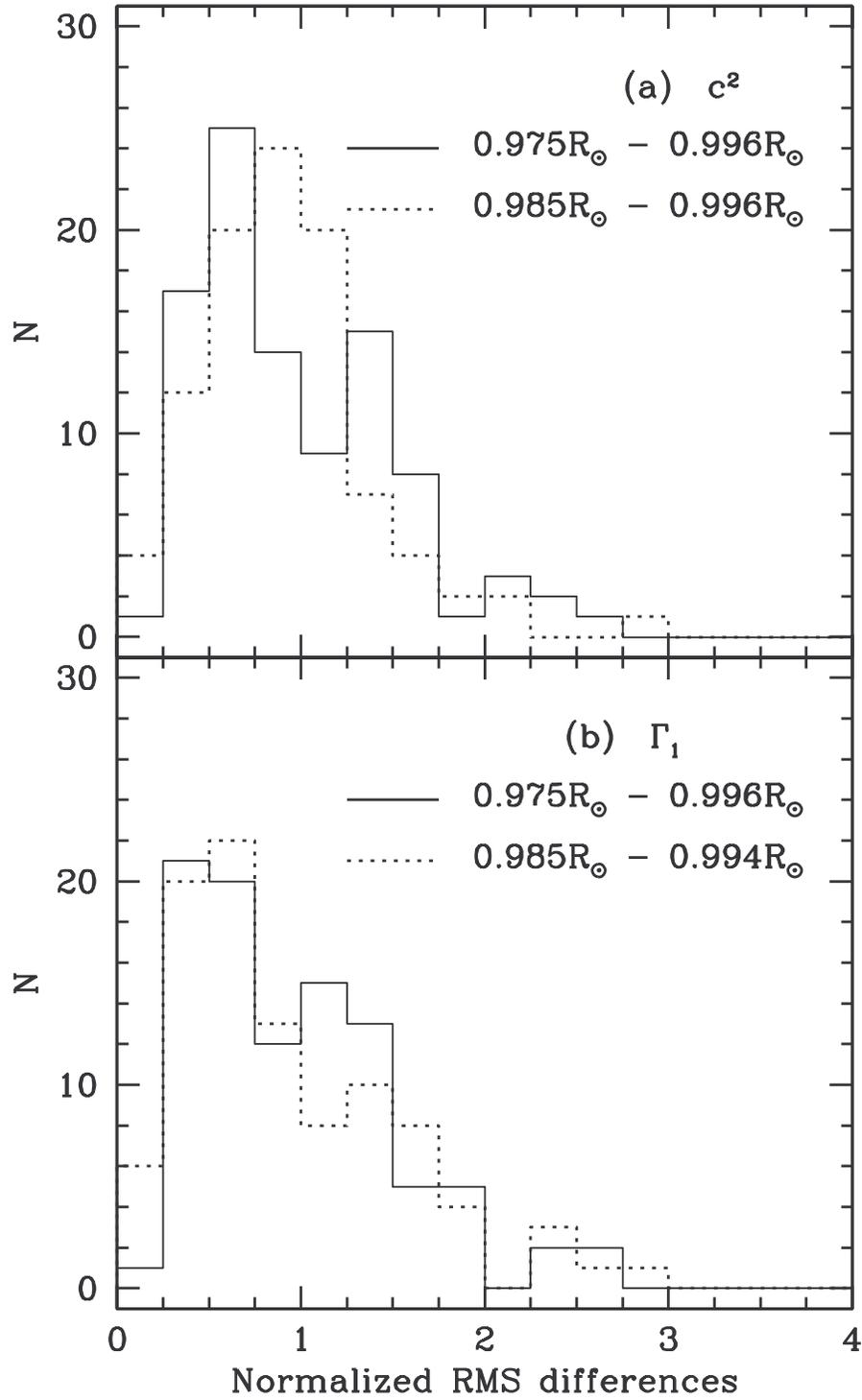}
\caption{Histogram of the RMS differences between SOLA and RLS inversion
results normalized to the errors. Panel (a) shows $c^2$ results and Panel (b)
the $\Gamma_1$ results.}
\label{fig:histo}
\end{figure}

\begin{figure}
\plotone{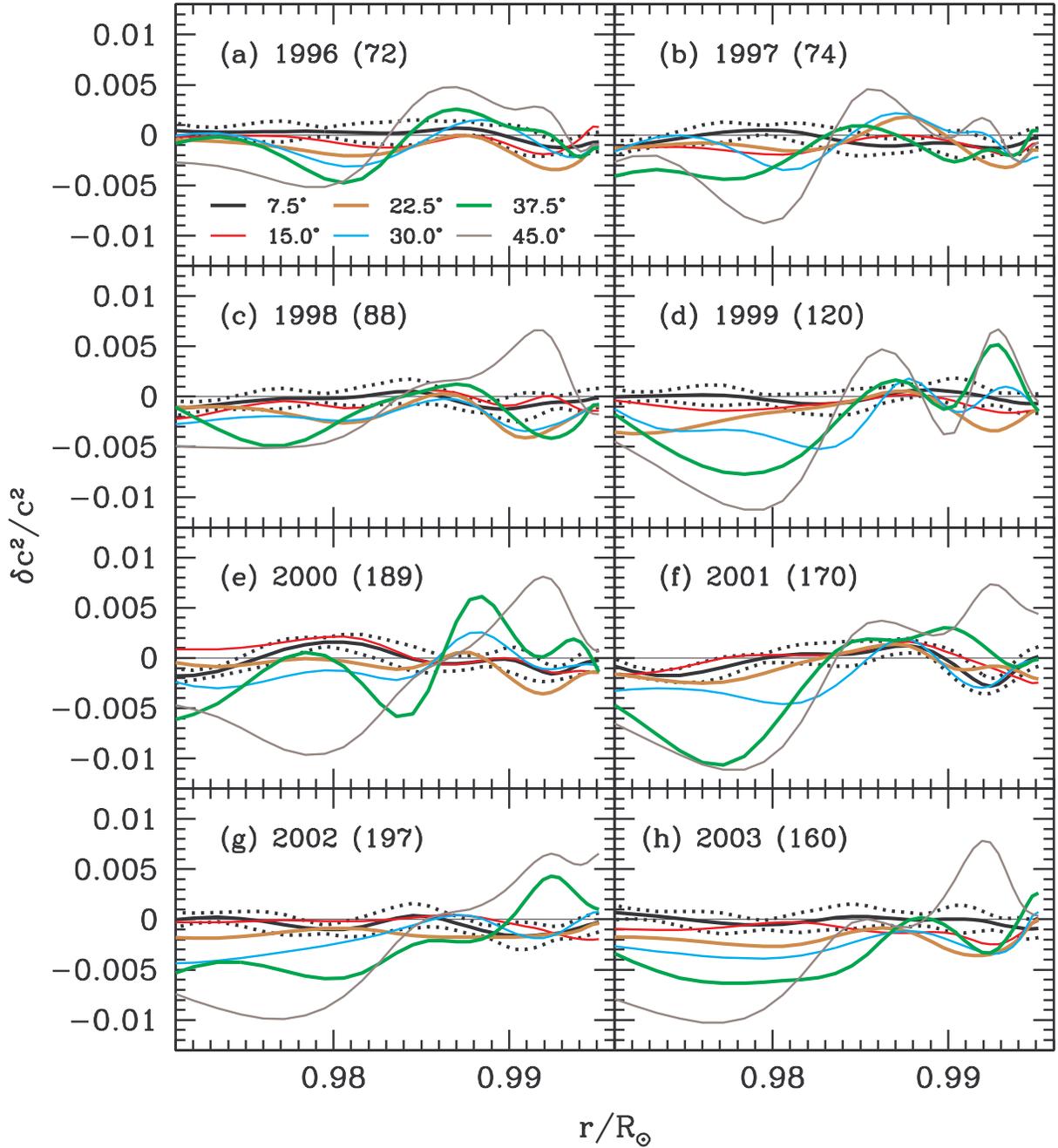}
\caption{The north-south averaged relative differences in the squared sound
speed $c^2$  between the higher latitudes and the solar equator, plotted
as a function of depth for the eight periods analyzed.  Only RLS results
are shown; SOLA results are similar. The dotted lines
are $1\sigma$ errors. For the sake of clarity, the errors are shown only for the
latitude of $7.5^\circ$. All differences are in the sense (Higher latitude $-$ Equator).
The panels are labeled by the calendar year of the data sets, the numbers in
parentheses being the 10.7 cm radio flux in units of SFU for the corresponding
period.} 
\label{fig:csq}
\end{figure}

\begin{figure}
\plotone{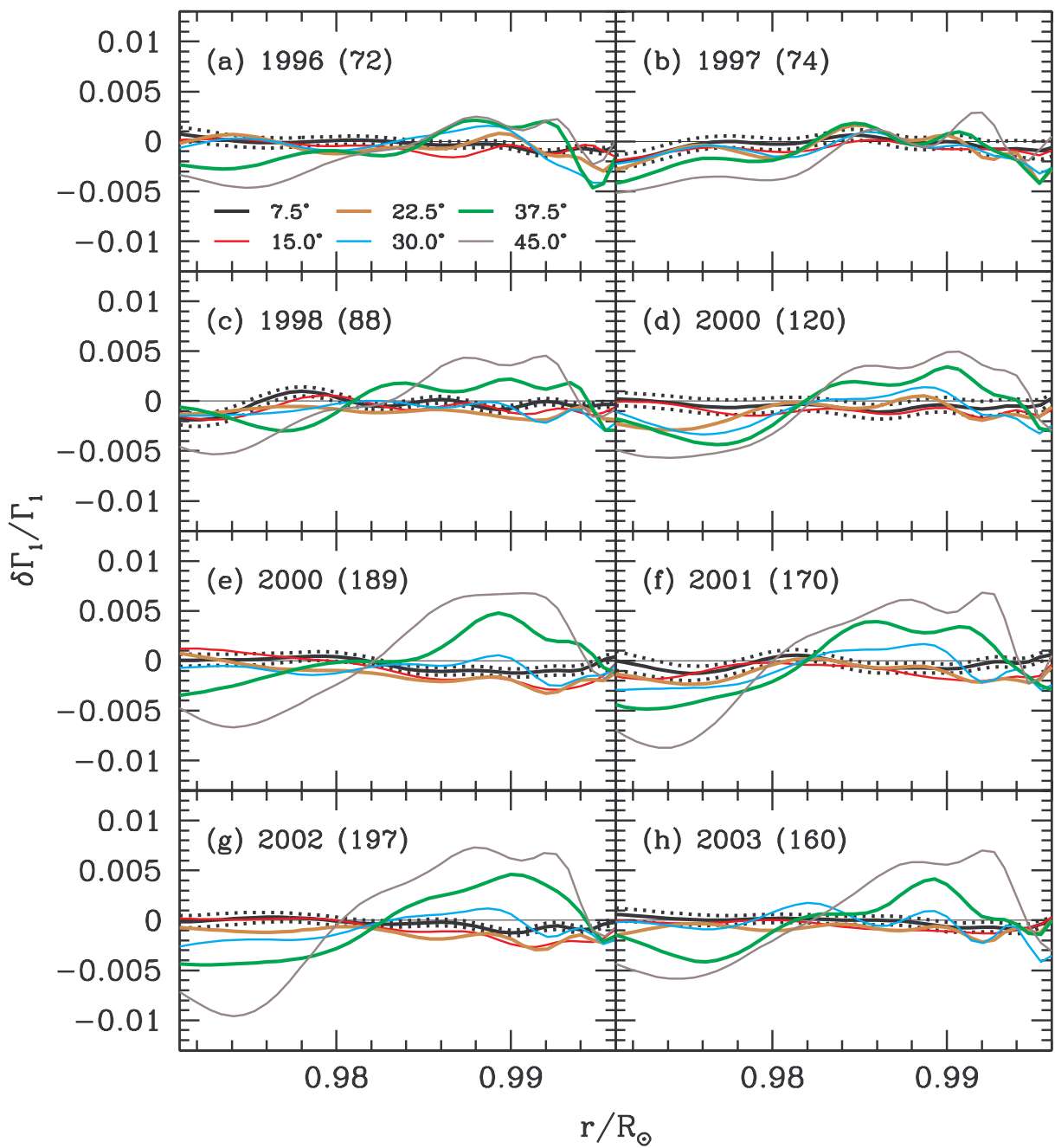}
\caption{The north-south averaged relative differences in the adiabatic
index $\Gamma_1$ between the
higher latitude and the solar equator as a function of depth for different
analysis sets.  The same remarks about error bars and labels for
Fig.~\ref{fig:csq} apply.}
\label{fig:gam}
\end{figure}

\begin{figure}
\plotone{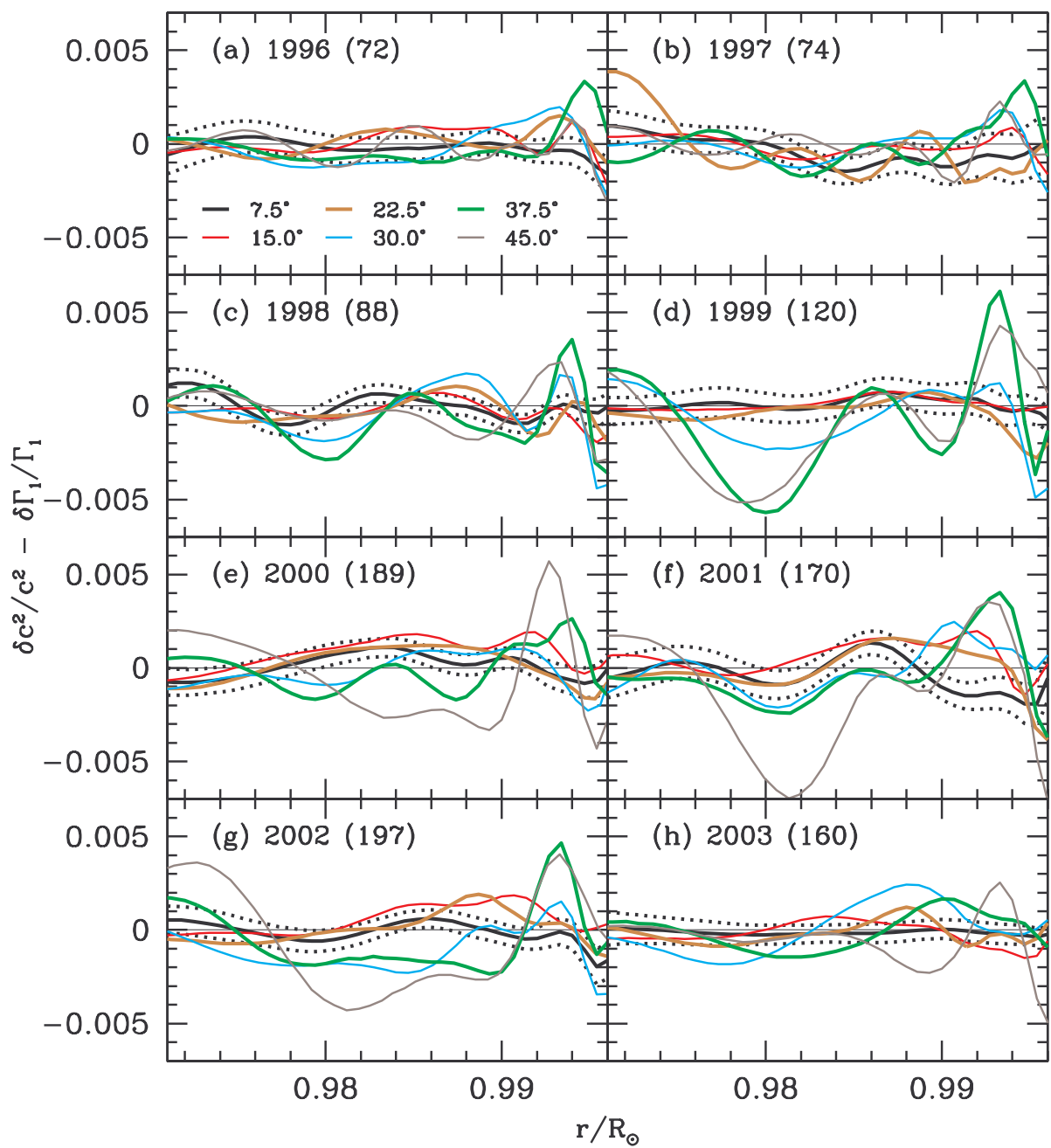}
\caption{The north-south averaged relative differences in the inferred
thermal parameter $c^2/\Gamma_1$ ($P/\rho$).
as a function of depth.  The error bars and labels are as described in
Fig.~\ref{fig:csq}.}
\label{fig:dif}
\end{figure}

\begin{figure}
\plotone{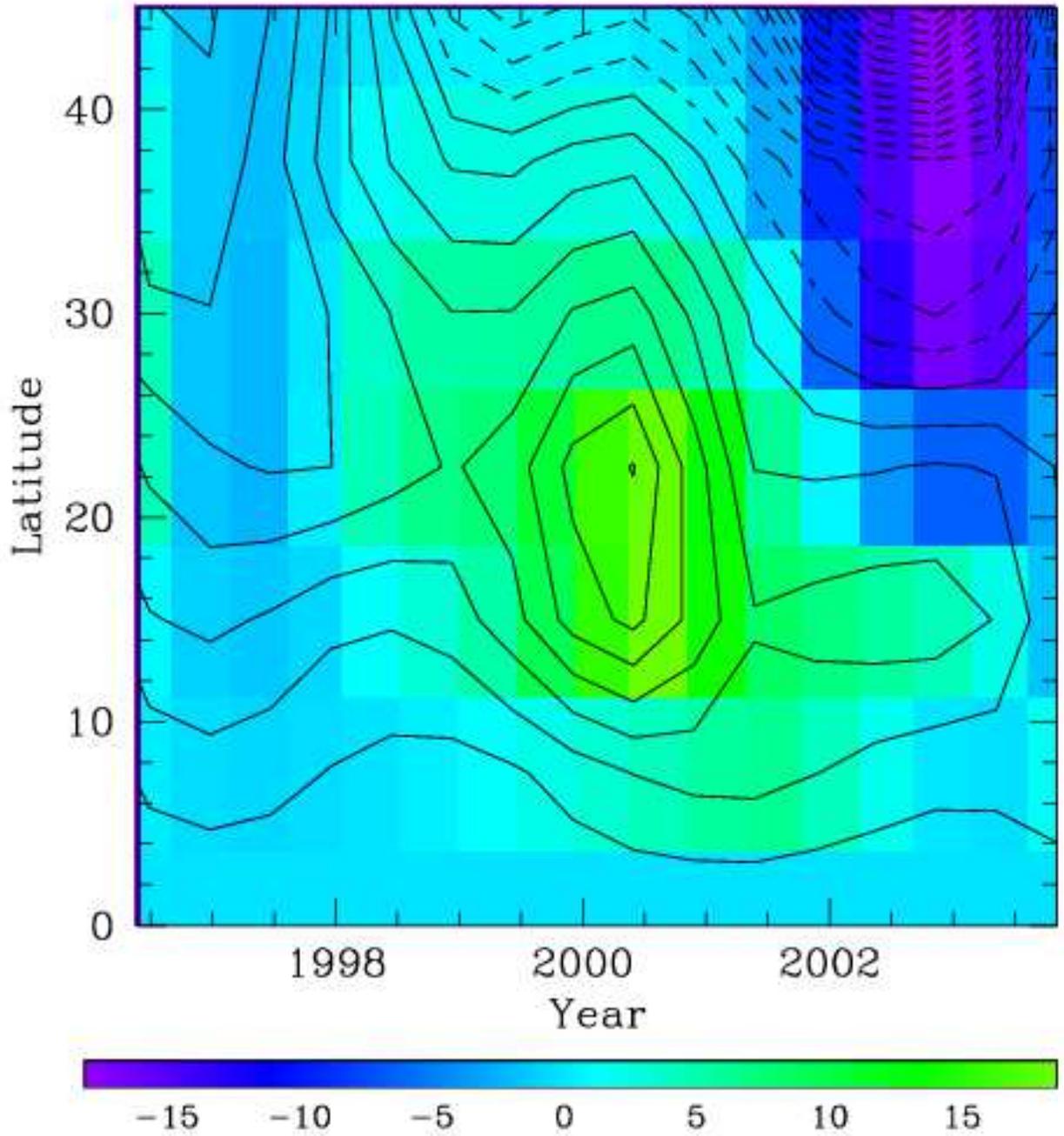}
\caption{The surface term from the inversions (contours) overplotted on the
MAI difference between the higher latitudes and the equator (color image).
The MAI difference is in units of Gauss. The surface term units are arbitrary,
with solid contours showing positive values and dashed contours showing
negative values.}
\label{fig:mai}
\end{figure}

\begin{figure}
\plotone{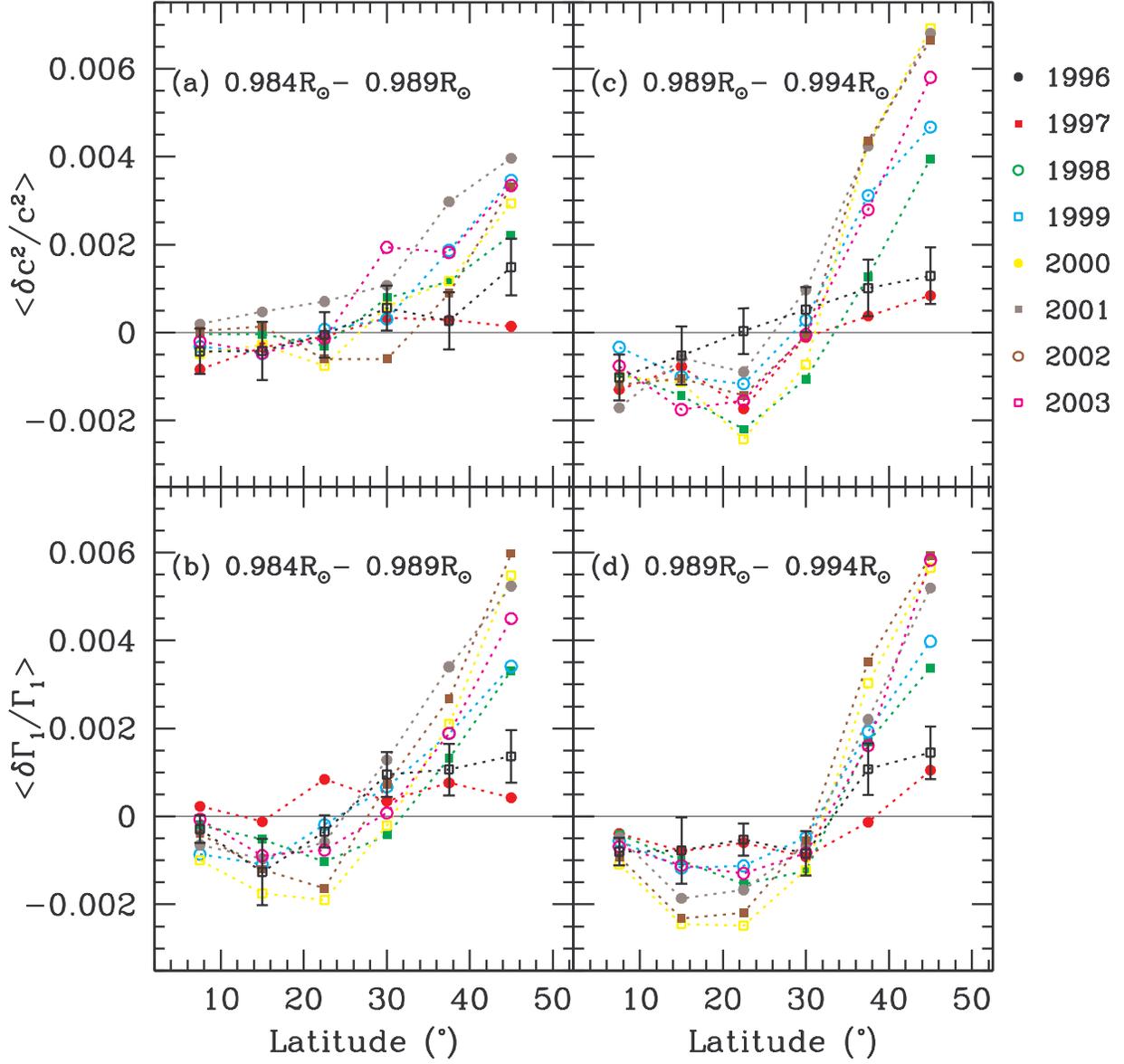}
\caption{The averaged sound-speed and $\Gamma_1$ differences plotted as a function of
latitude. Panels (a)--(b) are the averages over the depth range $0.984$--$0.989 R_\odot$,
while panels (c)--(d) are averages over the range $0.989$--$0.994 R_\odot$.
Error bars are shown for the 1996 results; they are comparable in other years.
The points are connected to guide the eye.
}
\label{fig:lat}
\end{figure}

\begin{figure}
\plotone{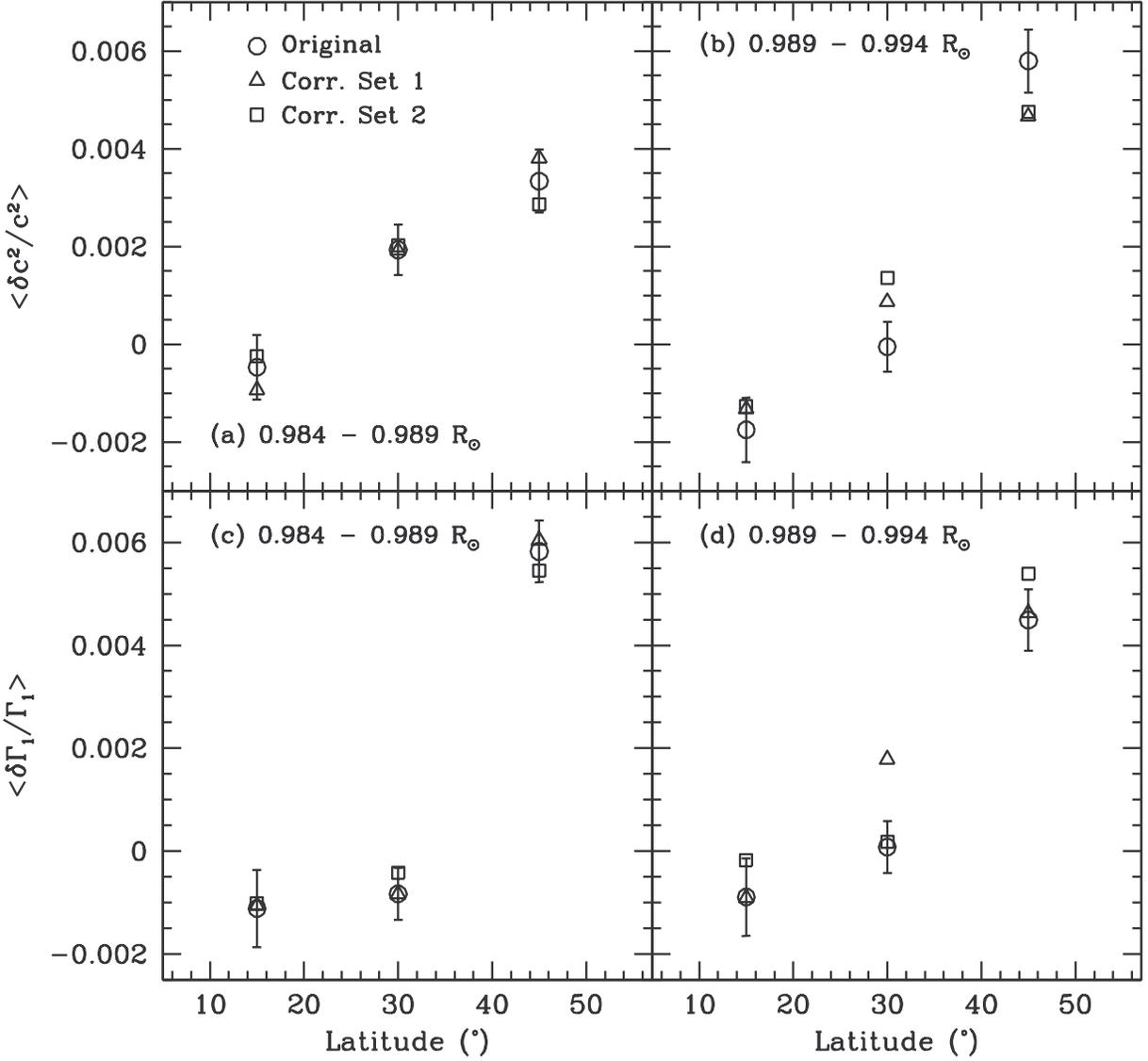}
\caption{A comparison of the results obtained from the year 2003 (CR 2009)
data and the data corrected for possible projection effects.
Corr.~Set 1 refers to corrections using
equatorial data from 1996 and Corr.~Set 2 are corrected using 1998 data.
The error bars represent the $1\sigma$ statistical errors for the
original data set. The errors for the corrected sets are roughly a
factor of 1.4 larger.
}
\label{fig:compcor}
\end{figure}

Since we find fairly substantial latitudinal effects, it is important to
look at possible systematic errors in the results, particularly those
due to errors in frequencies of higher-latitude regions caused by
projection effects.
Although a comparison between Fig.~\ref{fig:freqdif} and
Fig.~\ref{fig:freqcor} suggests that it is unlikely that the latitudinal
differences we see are a result of 
projection effects, nonetheless we need to test this. To do so,
we have subtracted the frequency differences shown in Fig.~\ref{fig:freqcor}
from the frequency differences that we inverted. We then
inverted these ``corrected'' sets of frequency differences.
We average the results within
the same radius ranges as the results in Fig.~\ref{fig:lat}. The comparison for one
epoch is shown in Fig.~\ref{fig:compcor}.
In the inner radius range, the results of the corrected and uncorrected sets
are very 
similar, all three sets being within $1\sigma$ of each other,
where the standard deviations refer to those of the original set.
(The errors in the ``corrected'' data sets are larger since the
correction involved subtracting out fitted frequencies which themselves
have errors.)
The results for the outer radius range show greater differences.
The differences are more
than $1\sigma$ of the original results, but if the errors in the
corrected set are considered, the results are within $1.5\sigma$.
Because of this small difference, we are comfortable with the
results of the original data, and we did not try to correct
all of the data sets. Our decision is also supported by the fact that the
results for the two corrected sets do not always lie on the same side
of the original result, so they do not appear to be
systematic. We conclude that
Fig.~\ref{fig:lat} does represent the difference in structure between
the higher latitudes and the equator.

The latitudinal variation of sound speed is even seen for the low-activity
epochs, as was seen in the global-mode work of Antia {\it et al.}~(2001).  This 
may seem surprising if one assumes that the asphericity is completely a result of 
the magnetic field distribution. However, the latitudinal structure
difference can depend on other factors. A latitudinal temperature 
difference is expected even in the absence of magnetic fields
because of solar differential rotation which can produce a relatively cool
equator and warmer poles (Thompson {\it et al.}~2003). Such
a temperature differential in the outer layers translates into a difference in sound speed
and $\Gamma_1$.
Numerical simulations also show
latitudinal temperature variations (Brun \& Toomre 2002; Brun {\it et al.}~2004, Miesch {\it et al.}~2006, etc.),
however, none of the simulations go to shallow enough layers to make a
direct comparison with our results. It should be noted, however, that the
simulations do not completely agree with the global-mode results.
The global-mode results show a cool equator and warmer higher latitude
region, while the outer layers of the simulations show a warmer equator and
pole with cooler regions in between. The deeper layers in the
simulations do show a cooler equator and warmer higher latitudes.

\subsubsection{Changes with time}

\begin{figure}
\plotone{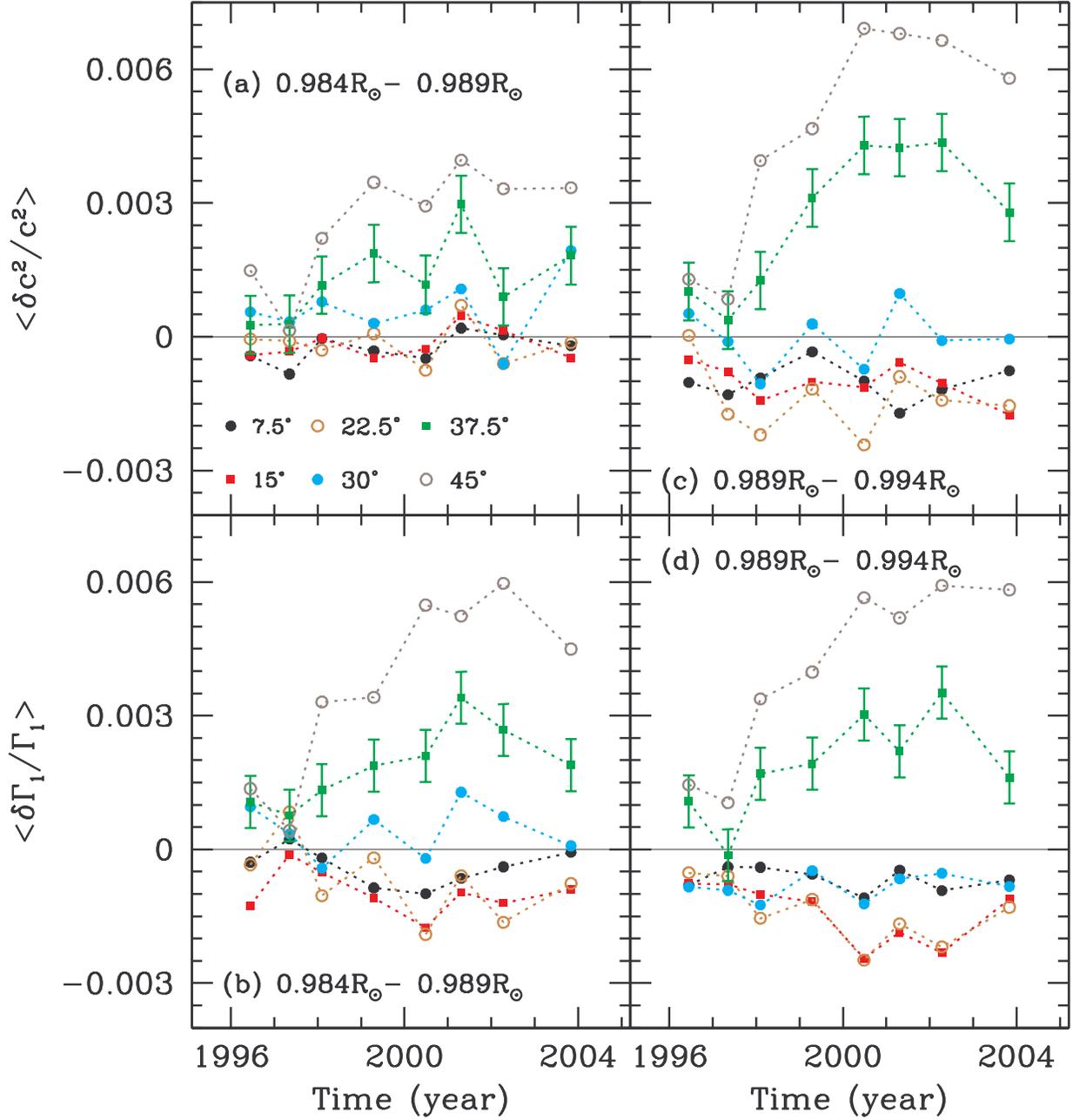}
\caption{The averaged sound-speed and $\Gamma_1$ differences plotted as a function of
time. Error-bars are only plotted on one set for the sake of clarity.}
\label{fig:time}
\end{figure}

In order to look in detail at the time dependence of the latitudinal
distribution of structure, we plot the averaged results as a function of time for
different latitudes. The results are shown in Fig.~\ref{fig:time}.  This figure
makes it clear that the differences in structure between the highest latitudes and the
equator show the greatest changes with time. This is not surprising,
since the difference in magnetic field between the equator and the
highest latitudes has been steadily increasing over the course of the cycle. 
This can be seen from the underlying color image in Fig.~\ref{fig:mai}. The changes are
generally larger over the outer radius range than the inner one.
This is particularly true for the sound speed differences. 
Again, for $\Gamma_1$ we can see the difference in behavior
below the latitude of $30^\circ$ and above it, with $\delta\Gamma_1/\Gamma_1$ 
being consistently negative for lower latitudes and positive for higher ones.
This a reflection of the fact that latitudes below about $30^\circ$ have
larger values of the MAI than the equator, while latitudes above $30^\circ$
have smaller values.

At lower latitudes, the sound speed differences show much smaller
changes with time, though again the changes are larger in the outer
radius range.  $\Gamma_1$ shows a similar variation. It could be asked
why the lower-latitude  regions  shows such a small change with time, 
when they are where the greatest concentration of magnetic
fields lie. There are two possible reasons for this. The first is
the rather poor latitudinal resolution of this work: the equatorial region
spans a latitude range of $\pm 7.5^\circ$, similarly the zone
centered at $7.5^\circ$ spans the region from the equator to $15^\circ$, the $15^\circ$
strip is from $7.5^\circ$ to $22.5^\circ$, {\it etc.} As a result, even if
there are large changes at $15^\circ$ with respect to the equator, the poor
resolution will reduce the apparent changes. Also, because of the
large latitudinal range covered by the equatorial strip, substantial magnetic
activity  changes take place within it. Since all of our work is
differential, these change are subtracted from those at other latitudes. Since the 
activity change in the lower latitudes has the same sign as the equator, the
change in the differences between these regions and the equator is smaller. For the
higher latitudes, on the other hand, any changes in structure with respect to the
equator are amplified, since the magnetic field strength changes in the opposite
direction.

\subsubsection{Dependence on activity}

Previous work by Basu {\it et al.}~(2004) has shown that the sound-speed
and $\Gamma_1$ differences between two regions increase if the
difference of the MAI's of the two regions increase. Since the MAI differences
between the equator and the other latitudes change with time, we have
investigated the correlations between MAI difference and the sound-speed
and $\Gamma_1$ differences.
We find that the linear
correlation coefficient between the $\Gamma_1$ differences
and the MAI differences for the 0.989--0.994$R_\odot$ radius range is only $-0.30$ when
all data sets are used. However, the correlation rises to 
$-0.56$ if only sets with $|\delta {\rm MAI}|> 5$~G are included.
For sound speed, the correlation coefficients are smaller ($-0.27$ and $-0.46$
respectively).
This small correlation  is in seeming contradiction to the results of Basu {\it et al.}~(2004),
but not if we take into account the fact that in that paper we were dealing
with MAI differences that were much larger (up to a factor of 16) than those in
this work, and Basu et al~(2004) had seen that the sound-speed and
$\Gamma_1$ differences do not change much for $|\delta {\rm MAI}|$ less than
about 30 G, and we are dealing with much smaller differences here.

\begin{figure}
\plotone{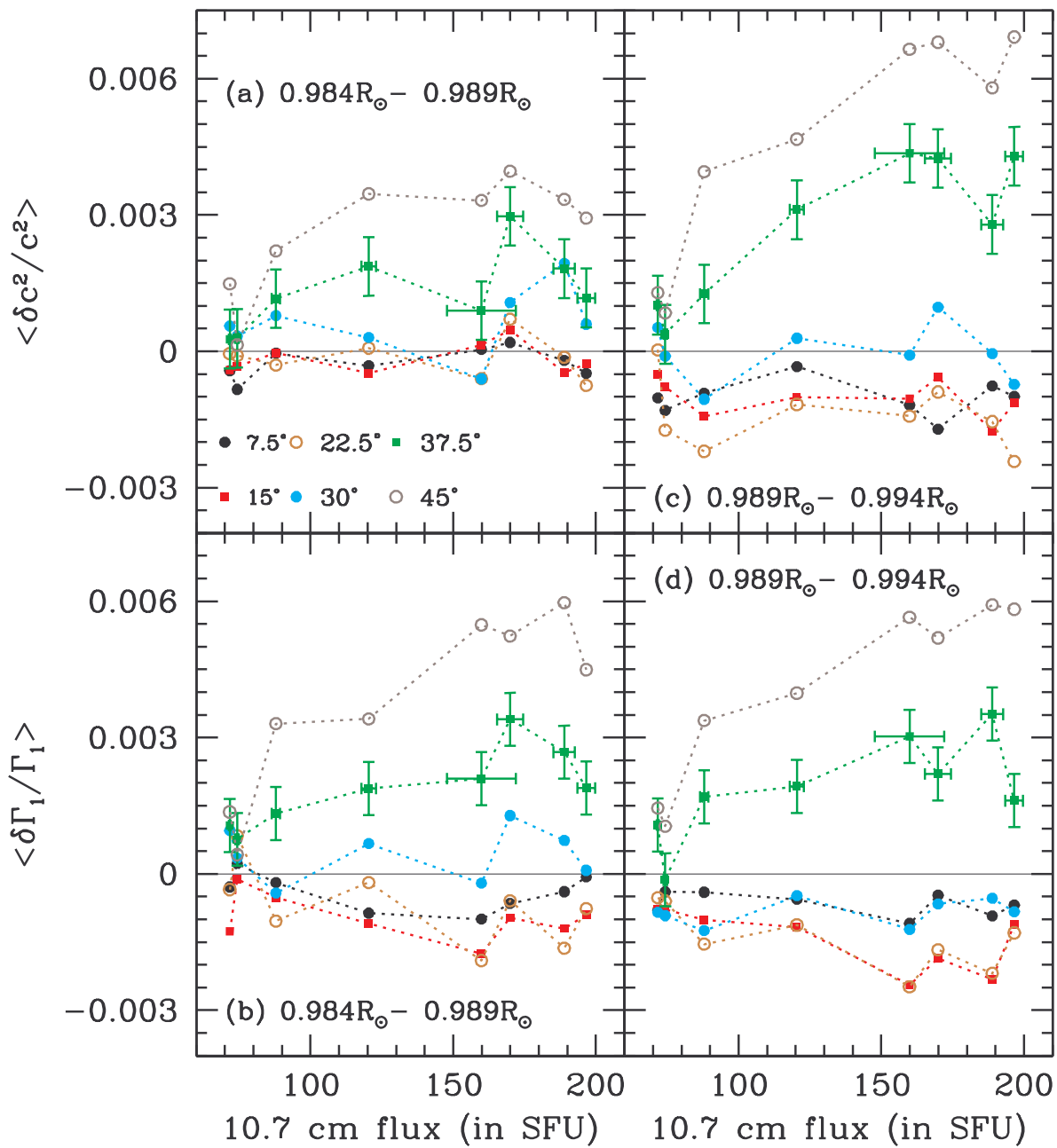}
\caption{Same as Fig.~\ref{fig:time}, but plotted as a function of the
10.7 cm flux, a measure of global solar activity. The flux is in
solar flux units, $10^{-22}$ J s$^{-1}$ m$^{-2}$ Hz$^{-1}$.
}
\label{fig:act}
\end{figure}

In order to study the dependence of the results on the level of
global solar activity, we have plotted them as a function of the
10.7 cm radio flux.
The results are shown in Fig.~\ref{fig:act}.
We find  as before that different latitudes show different trends.

We find that the differences between the different latitudes and the
equator change with activity, particularly at the highest latitudes studied,
where both $c^2$ and $\Gamma_1$ increase with activity.
There seems to be a saturation of the increase at very high activity
levels. This effect was also seen when we studied sound-speed
differences between active and quiet regions (Basu {\it et al.}~2004). They
found that the differences became nearly constant above a certain level of
magnetic field strength. During the late part of the solar cycle covered
by CR 1988 and CR 2009 (2002 and 2003), the average magnetic field
at the equator is significantly
larger than that at higher latitudes; that would also contribute to
some of the observed differences.
At lower latitudes, sound speed and $\Gamma_1$ appear to decrease
with increased  activity, but the results are not
statistically very significant.

\begin{figure}  
\plotone{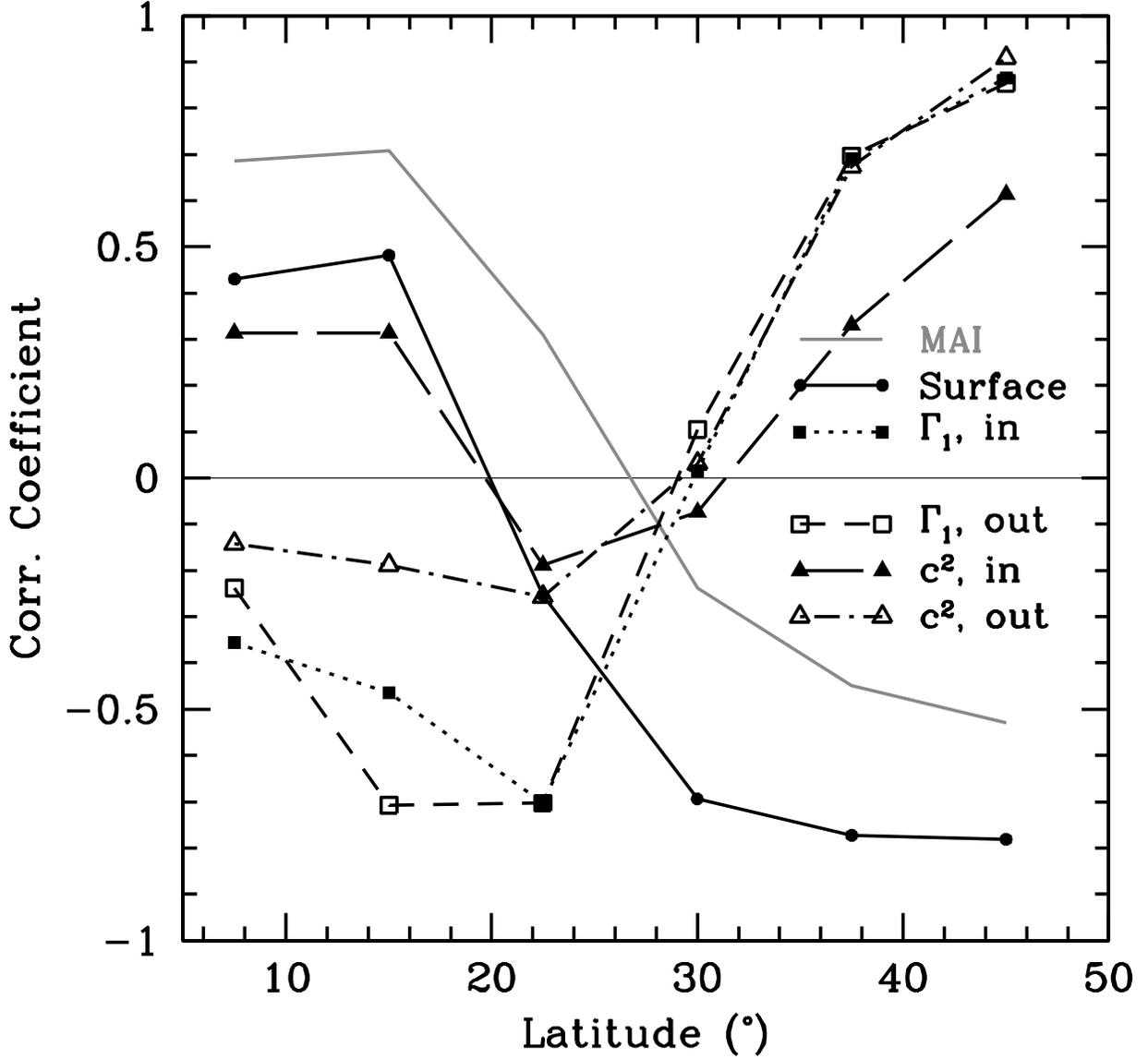}
\caption{The correlation coefficient between different quantities and
the 10.7 cm radio flux plotted as a function of latitude. The gray line shows
the correlation of the 10.7 cm flux with the MAI difference between a given latitude 
and the equator. $\Gamma_1$
refers to the relative differences of $\Gamma_1$ between the higher latitudes and the
equator, while $c^2$ denote the relative differences of $c^2$. The terms `in' and
`out' refer to the inner and outer radius ranges, i.e., $0.984$--$0.989 R_\odot$ 
and $0.989$--$0.994 R_\odot$ respectively.}
\label{fig:corr}
\end{figure}

We quantify the change of $\delta c^2/c^2$ and $\delta\Gamma_1/\Gamma_1$ with
the 10.7 cm flux by calculating the linear correlation coefficient. 
The correlation of the surface term, and $c^2$ and $\Gamma_1$ differences at
each latitude with the 10.7 cm flux is shown in Fig.~\ref{fig:corr}. We can
see a positive correlation at high latitudes and a largely negative correlation at
lower latitudes. 
We find that
the change with solar activity is stronger for the high latitude regions than
at low latitude regions, and the correlation with the 10.7 cm flux is stronger 
for $\Gamma_1$ than for $c^2$, and the outer-radius range (0.989--0.994$R_\odot$)
shows a more pronounced change than the inner radius range (0.984--0.989$R_\odot$).
We find that above $30^\circ$, the $\Gamma_1$ variation has an average
correlation coefficient of
$0.78$, while below $30^\circ$ it is $-0.5$. At $30^\circ$, the results show no
correlation with activity at all. The correlation for $c^2$ differences is
much weaker: for latitudes greater than $30^\circ$ it is 0.47 and 0.79
for the inner and outer radius ranges respectively.
The lower-latitude range shows correlations only of
0.14 and $-0.19$, which are probably not statistically significant.
The difference in behavior between the low and high latitudes can
perhaps be attributed to the different contributions of the fields
near the equator to the differences. The correlation with
MAI difference also changes sign around the same latitude $30^\circ$. This could
be because of the progression of activity toward lower latitudes with the cycle.
As the activity shifts towards the equator towards the end of the solar cycle,
the MAI difference between high latitudes and the equator changes sign.

\subsection{Absolute time differences}
\label{subsec:time}

\begin{figure}
\plotone{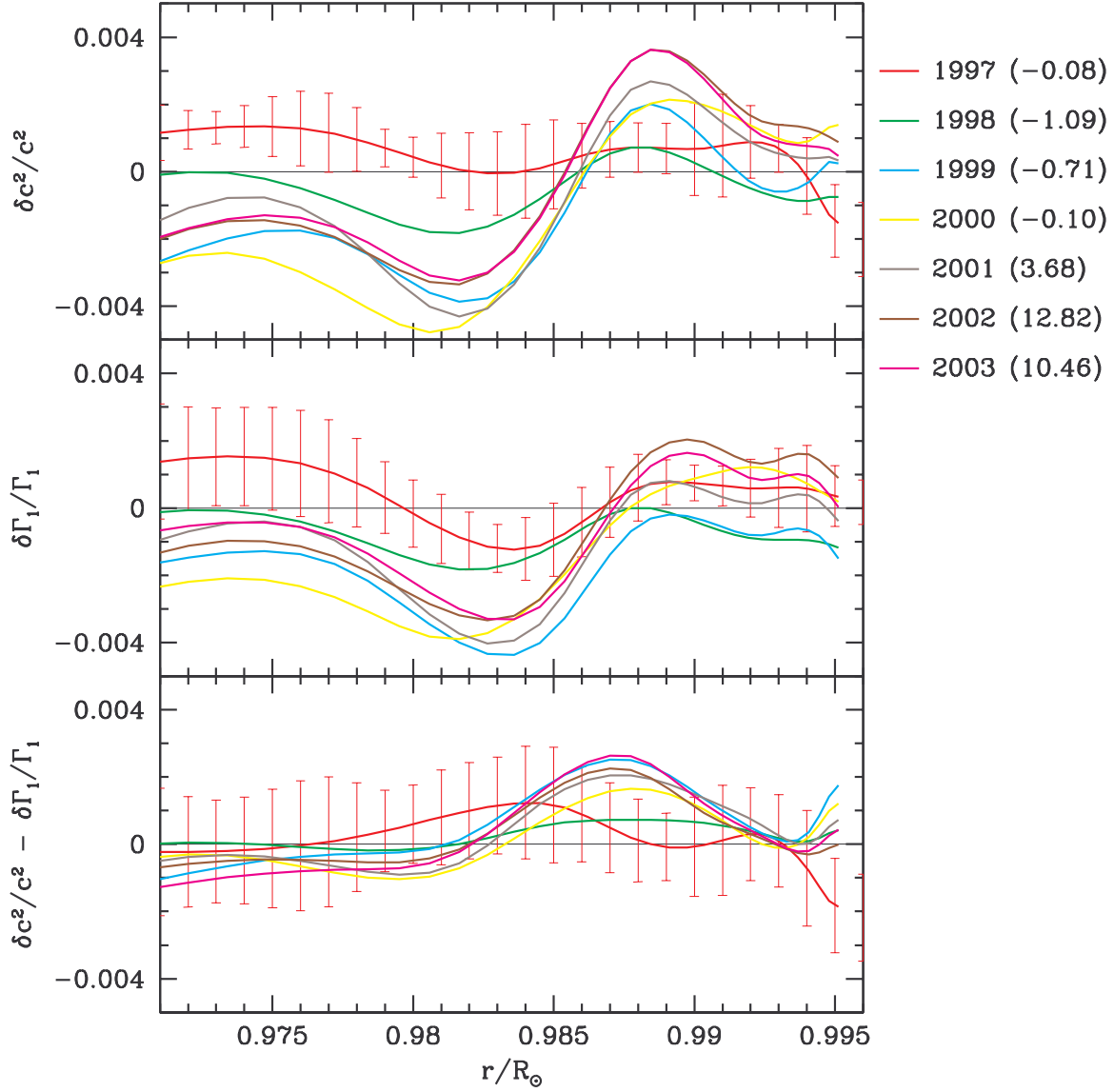}
\caption{The sound speed and $\Gamma_1$ differences at the solar equator. The
different lines show the differences between different epochs and 
CR 1910 plotted as a function of depth. Only the RLS results
are shown; the
SOLA results are similar. Error bars are shown for only one comparison year
(1997) for the sake of clarity.
The differences are in the sense (Later epoch $-$ 1996). The numbers in
brackets denote the MAI differences (in Gauss) at the equator between the
later epochs and 1996.
}
\label{fig:csqtime}
\end{figure}

We have so far focused on the latitudinal distributions of $c^2$ and
$\Gamma_1$ during each of the several Carrington rotations analyzed.
The use of contemporaneous data minimizes systematic frequency errors
caused by year-to-year variations of the MDI instrument, such as plate
scale.  If we 
disregard this potential problem, we can estimate the real change 
in $c^2$ and $\Gamma_1$ in the Sun as a function of time, and not
just the change between different latitudes and the equator.
Figure~\ref{fig:csqtime} shows how the sound speed and $\Gamma_1$
profiles change with time at the equator.
We can see that there are changes that are statistically significant
when the MAI differences between the equator at various epochs are
not too small.

\begin{figure}
\plotone{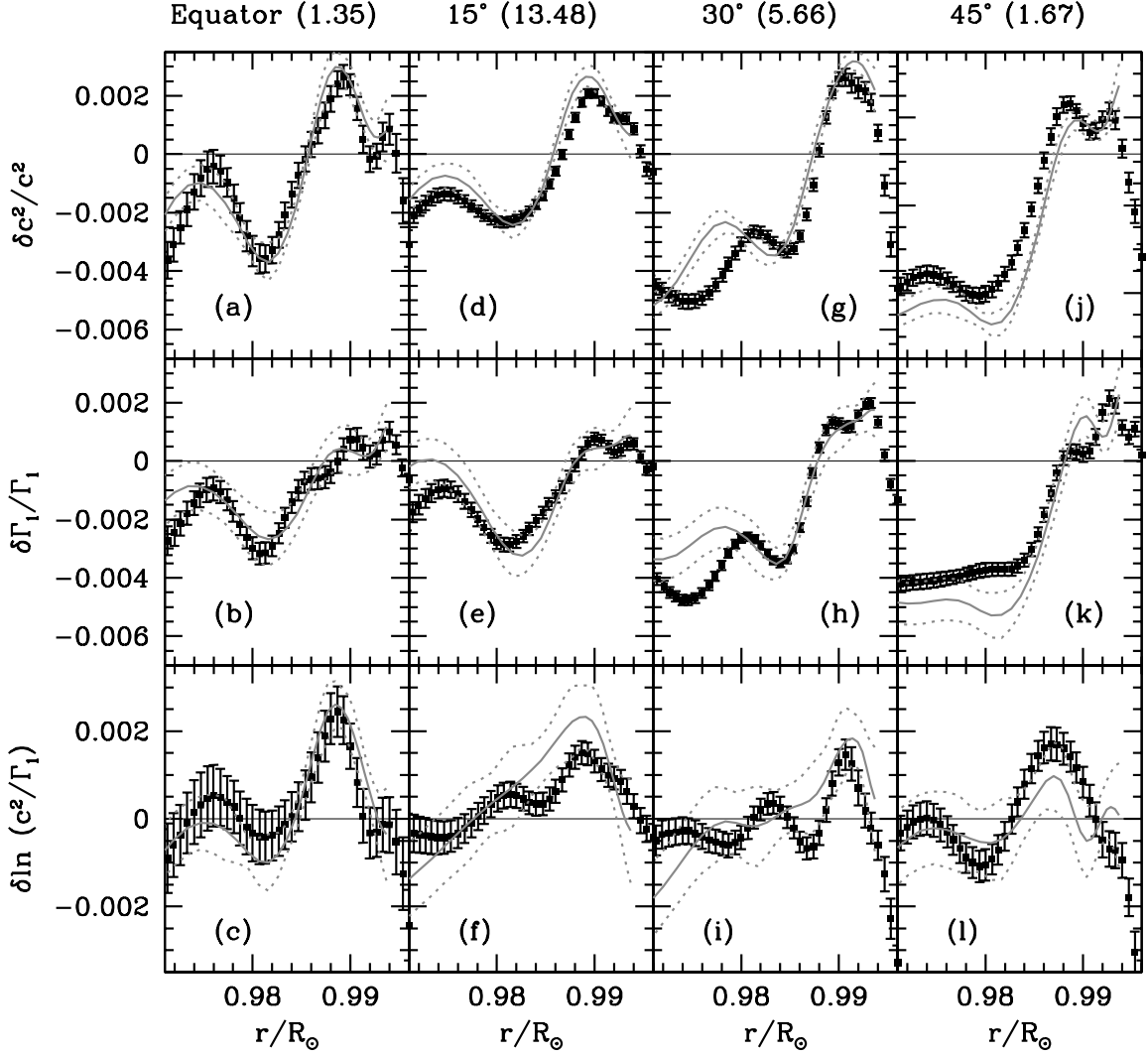}
\caption{The sound-speed, $\Gamma_1$ and $c^2/\Gamma_1$ differences between the high and
low activity sets at different latitudes. The latitudes are
marked at the top with the MAI difference in units of Gauss given
in brackets. The differences are in the sense 
(High Activity $-$ Low Activity). 
The black points are SOLA results, the error
bars representing 1$\sigma$ 
errors, and the gray continuous lines are RLS results with the gray  dotted lines
showing the 1$\sigma$ error limits.  
}
\label{fig:high_low}
\end{figure}

Since the errors in the results shown in
Fig.~\ref{fig:csqtime} are large, it is difficult to judge their
significance. In order to improve the signal-to-noise level of the results,
we averaged the power spectra for the three sets with lowest activity
(CR 1910, 1922 and 1932 in years 1996, 1997, and 1998 respectively)
to produce an average power spectrum for low activity times
We did the same for the three highest activity sets
(CR 1964, 1975 and 1988 in years 2000, 2001, and 2002).
These two average spectra were fitted in the same manner as the
other data sets. We then inverted the frequency differences between
the high- and low-activity sets at different latitudes.
The errors in the inversion results should improve by more than
the factor of $\sqrt{3}$ in the fitted frequencies themselves,
because the improved signal-to-noise ratio permits us to fit many more
modes. The sound-speed and $\Gamma_1$ differences between the high- and
low-activity sets are shown in Fig.~\ref{fig:high_low}. We note that
the fitting improvements have also translated into better agreement
between the RLS and SOLA inversion results.
The first thing to note from Fig.~\ref{fig:high_low} is that the differences
in $c^2$ and $\Gamma_1$ between times of low and high activity are
statistically significant over the whole depth range. The difference
for perturbations in $c^2/\Gamma_1$ is significant over a narrower range,
being essentially zero below about $0.98R_\odot$. The general shape of the
differences at the equator is similar to that seen in Fig.~\ref{fig:csqtime}.
For these sets we  find that the sound speed and adiabatic index in the near-surface
regions for the radius range $0.986 < r/R_\odot < 0.995$ was higher 
at all latitudes when the Sun was active.
This is the opposite of what we found when studying the difference
between active and quiet regions (Basu {\it et al.}~2004), where we found that
active regions had lower sound speeds in the near surface layers than the quiet
regions for $r/R_\odot  > 0.99$. The results are however consistent
for $r/R_\odot > 0.995$. There are several possible explanations
for this phenomenon. It could be that the expected average negative
sound-speed and $\Gamma_1$ differentials lie in a shallower radius range
because the MAI differences that we are dealing with in this work are smaller.
The results could also
mean that the frequency shifts for entire latitude zones of an
active Sun do not appear to be just the average of those for the active
regions. It is of course possible that some of this difference is due to
the fact that we are not including the direct effects of the magnetic
fields in inversions.  It is also possible that the results are affected
by time-dependent systematic effects, such as changes in the MDI focus.

\subsection{Differences between the northern and southern hemispheres}
\label{subsec:ns}

\begin{figure}
\plotone{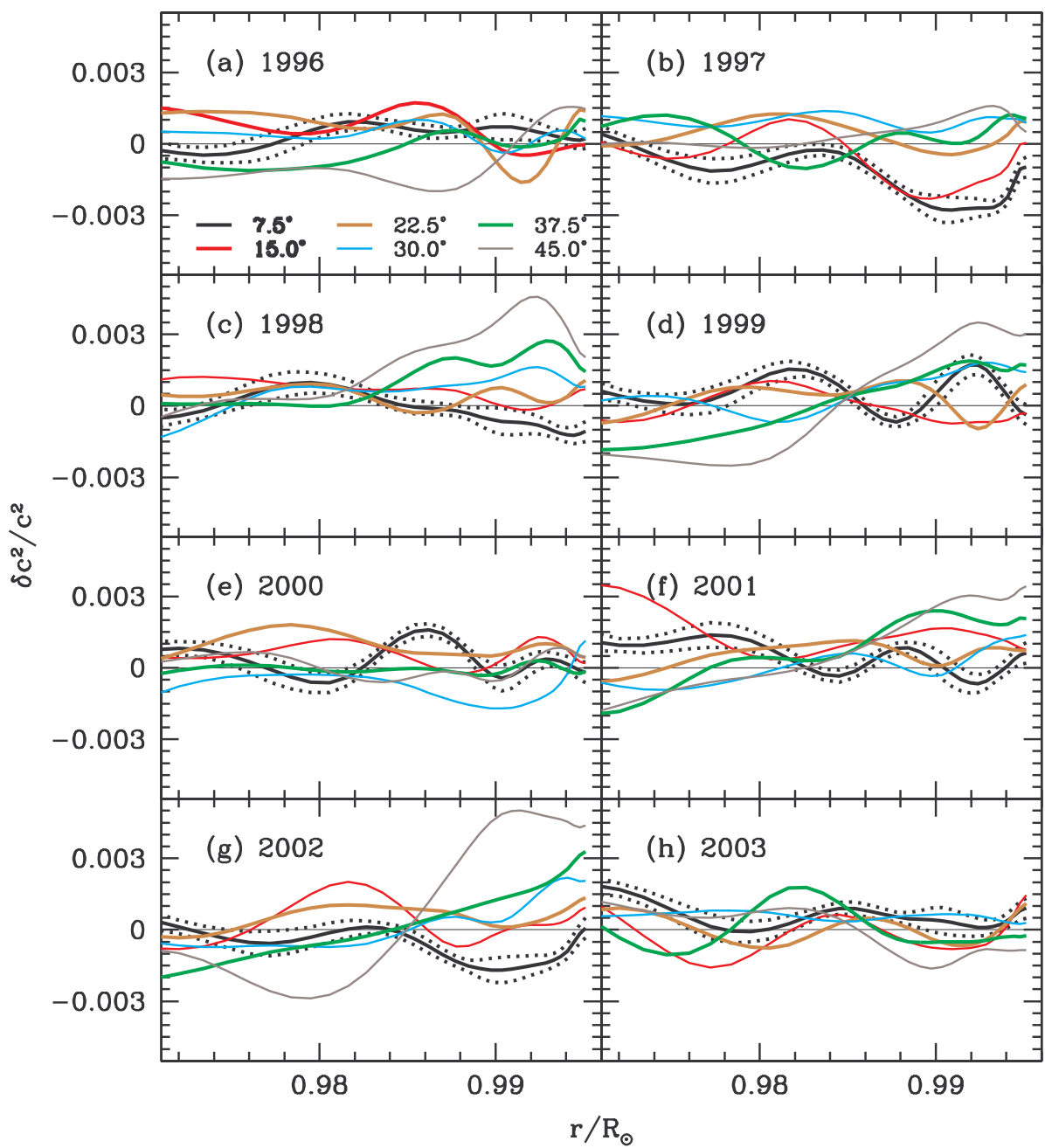}
\caption{The difference in sound speed between the northern and southern
hemispheres of the Sun as a function of depth. $1\sigma$ error limits are shown as
dotted lines. Errors are shown only for 7.5$^\circ$ for the sake of
clarity. Only RLS results are shown.}
\label{fig:csqnsdif}
\end{figure}

In the previous sections we examined only averaged results for symmetric
latitude zones in the northern and southern hemispheres. One of the
advantages of ring-diagram over global-mode analysis is that 
it permits us to look for north-south asymmetries by inverting the
frequency differences between corresponding latitude zones in the
two hemispheres.
Figure~\ref{fig:csqnsdif} shows the difference between the northern and southern hemisphere
sound speeds. The results for $\Gamma_1$, which are not shown, are quite similar.
There is no discernible pattern; there are clearly significant
hemispheric differences at some times for some latitudes, but no
consistent trends. We frequently find large asymmetries
at the higher latitudes ($37.5^\circ$ and $45^\circ$), but the
differences are not always statistically significant.

One of the more interesting features is the north-south asymmetry
at $\pm45^\circ$ latitude for data from 1999 and 2002 (see panels
{\it d} and {\it g} of Fig.~\ref{fig:csqnsdif}).  There is a 
very large difference between the northern and the southern hemisphere in the deeper layers.
We cannot be certain about the reality of that feature at $45^\circ.$
Haber {\it et al.}~(2002) found submerged counter-flowing meridional flow cells
at these depths in the northern hemisphere only in 1999 and 2001.
It has been argued that the the countercell feature in the
meridional flow can be caused by data errors (see e.g., Bogart \& Basu~2004);
if so, this feature in the sound-speed differences
could have the same origin. We do not see any similar feature in the 2001
data, however.  Haber {\it et al.}~(2002) 
did not analyze 2002 data, so the present result cannot yet be directly
compared with meridional flow asymmetries.

\begin{figure}
\plotone{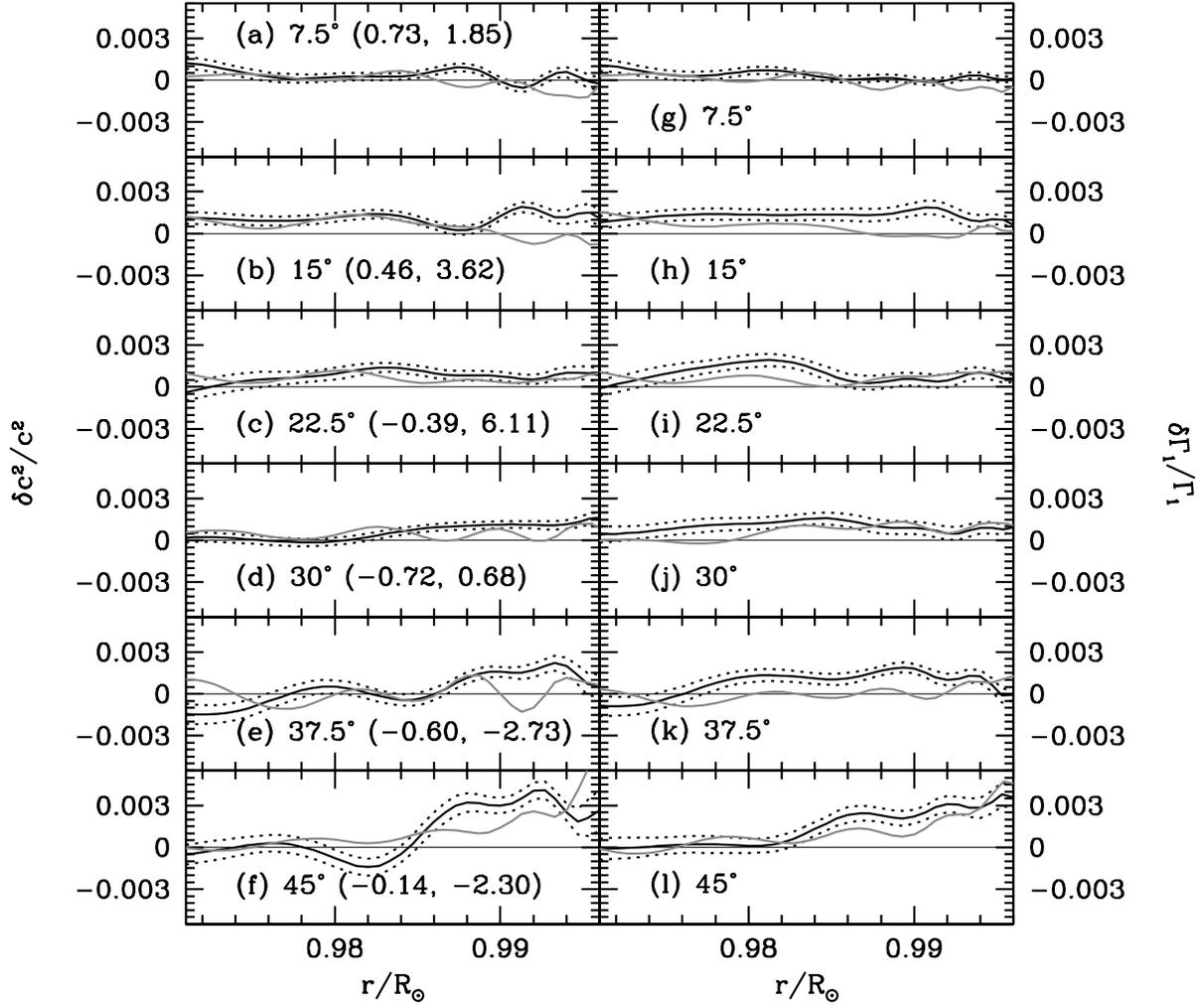}
\caption{The north-south differences in $c^2$ (panels a--f) and $\Gamma_1$ (panels
g--l) for the averaged high activity set (black) and the averaged low activity set
(gray). Differences are in the sense {(North $-$ South)}. The bracketed pair
of numbers in panels a--f are MAI differences for that latitude for the
low- and high-activity sets respectively.
Error (as dotted lines)  are shown for only the high-activity set for clarity. Only SOLA
results are shown; RLS results are very similar.}
\label{fig:av_nsdif}
\end{figure}

Given that the north-south asymmetries for individual epochs are small,
we again use the
low-activity and high-activity average spectra of \S~\ref{subsec:time} to improve
the signal-to-noise ratio. We examine the difference signals
separately for the low-activity and high-activity data. 
The results for
both $c^2$ and $\Gamma_1$ differences between the northern and the southern
hemispheres are shown in Fig.~\ref{fig:av_nsdif}. 
It is quite clear that there is little evidence of asymmetry at low
latitudes, as the differences
are generally at the 2--3$\sigma$ level. 
At higher latitudes
($> 30^\circ$), the  difference between the hemispheres is clearer,
particularly for the high-activity set.
In general, the asymmetry appears 
larger  at the higher activity epoch. 
It thus appears that there was some significant difference
in structure at high latitudes during the active phase of cycle 23.
This by itself is not surprising: many activity-related observations
show north-south asymmetry; e.g., de Toma {\it et al.}~(2000) found that
at the end of cycle 22 and onset of cycle 23, Kitt Peak magnetograms show a
north-south asymmetry. North-south differences have also been found
in the distribution of solar flares (e.g., Bai 1990) and magnetic filaments
(e.g., Duchlev \& Dermendjiev 1996), and in the photospheric magnetic flux
(e.g., Howard 1974; Knaack {\it et al.}~2004). 
The fact that the low-activity set has a lower north-south difference
than the high-activity one is consistent with the fact that the MAI differences
for the low-activity sets are smaller than those of the high-activity ones.
However, for any given set, the differences are not what we expect from the
MAI differences. We would have expected, for example, a larger difference at
22.5$^\circ$ for the high-activity set than for the 45$^\circ$ set, but we
see the opposite.
These results could again be an indication that
averaging of non-contemporaneous ring-diagram spectra may not be the correct
thing to do because of possible time-dependent systematic observational effects.

\section{Conclusions}
\label{sec:conclu}

We have analyzed data from eight full Carrington rotations to determine
the latitudinal distribution of the axisymmetric components of sound speed
and $\Gamma_1$, and to determine whether
those distributions change with time and magnetic activity level.
Ring-diagram analysis allows us to determine frequencies
of high-degree modes, permitting us to determine solar structure in
the near-surface layers, where most solar-cycle related changes are
believed to occur.  Our results are valid in the radius range
$0.975$--$0.996R_\odot$ and most reliable in the 
range $0.985$--$0.995R_\odot$. We have studied latitudes up to $\pm45^\circ$.

For all epochs studied, we find significant frequency differences
between the equator and the higher latitudes.  These differences cannot
be explained as being caused by projection effects alone; most of the
differences are  believed to reflect physical causes.
Assuming that the frequency differences are caused by differences in structure alone,
we have inverted them to determine the sound speed and the
adiabatic index differences between the equator and higher latitudes.
We find significant differences in structure between the equator and
the higher latitudes for all epochs.
The differences are largest for
latitudes $ > 30^\circ$. These regions tend to have higher sound speed and $\Gamma_1$ than 
the equator. The differences in $\Gamma_1$ between the equator and higher latitudes
imply temperature changes in these regions.  
Using global modes, Antia {\it et al.}~(2001) also found
a region of positive sound speed difference around a latitude of
$60^\circ$. Although the magnitude of the difference is much smaller
in deeper layers, of the order of $10^{-4}$, the qualitative
behavior is similar. 
We expect the relative magnitude of the
effect to increase in the near surface layers, as the global analysis
suggests that most of the temporal variations in the solar oscillation
frequencies arise from the surface layers. Because of the low pressure near
the surface, magnetic fields are expected to be more effective in
modifying the structure.

We find that the surface term from the inversions changes with time.
The surface term is reasonably correlated with the difference 
between the magnetic activity index of the higher latitude
regions and the equator. The correlation improves if only
$|\delta {\rm MAI} | > 5$~G regions are considered. The surface term
is also correlated with the 10.7 cm flux, though the correlation
is strongly dependent on the latitude. The latitudinal dependence
of the correlation coefficient can be explained by changes in
the latitudinal distribution of the magnetic field as solar activity
changes.

The asphericity of the Sun changes with time, with the difference
between the higher latitudes and the Equator increasing with increasing
levels of activity. The differences can be explained by the change 
in the magnetic field distribution on the Sun as the activity level
increases. The change with time and activity is in opposite
directions for latitudes above and below $30^\circ$.
This is consistent with the evolution of the magnetic field distribution.
As with the surface term, the differences are correlated both
with local and global magnetic field indices.
The magnitude of changes between the active and quiet periods seen in
this study is about a factor of 3--10 less than what was
observed between individual active and quiet regions (Basu {\it et al.}~2004).
This is expected, as the average magnetic field strength over a whole
latitude band is much smaller than that in active regions alone.
Although it is difficult to separate the effects of magnetic
field and structural variations, one may expect the magnetic field
to play a significant role near the surface. The ratio of magnetic
to gas pressure is expected to be comparable, to order of 
magnitude, to the variations found in sound speed.
At $r=0.99R_\odot$ a variation
of $10^{-3}$ may correspond to a magnetic field of 4000 G, which is
considerably larger than the typical magnetic field observed at the surface.

We have also looked at temporal variations of structure by inverting the
frequency differences between times when the Sun was active (2000--2002) and
when it was relatively quiet (1996--1998). Assuming  that there
are no time-dependent systematic errors in the data, these differences
were inverted to look at how the structure of the outer layers of the
Sun  differed during the two epochs. We find that there are significant
differences in the near-surface structure of the Sun during quiet and
active periods. These results, however, may be affected by systematic
observational variations.
It is also possible that the results are systematically
biased in some subtle way by the different projection effects involved
in tracking or not tracking the data.

We have also studied the north-south asymmetries in the structure of the
outer layers of the Sun. We find that when the Sun was active during cycle 23,
the sound speed in the northern hemisphere was higher than that in the
southern hemisphere, at least at high latitudes. Somewhat similar behavior is
seen for $\Gamma_1$. This can be explained by differing amounts of
asymmetry in the magnetic fields of the northern and southern hemispheres
when the Sun was quiet and when the Sun was active.

\acknowledgments
The authors thank the referee for constructive comments that have
resulted in a much better paper.
This work utilizes data from the Solar Oscillations
Investigation / Michelson Doppler Imager (SOI/MDI) on the Solar and
Heliospheric Observatory (SOHO). The MDI project is supported by NASA
grant NAG5-8878 to Stanford University. SOHO is a project of international
cooperation between ESA and NASA. This work was partially supported  by NASA
grants NAG5-10912 and NNG06GD13G, and NSF grant ATM 0348837 to SB.

\clearpage

\begin{deluxetable}{ccl}
\tablecaption{\em Data sets analyzed}
\tablehead{\colhead{Carrington}& \colhead{Dates} &\colhead{10.7 cm Flux}\\
\colhead{Rotation}&&\quad(SFU)$^1$\\}
\startdata
1910 & 1996:06:01 -- 1996:06:28 & $\phantom{0}71.7\pm 0.3$\\
1922 & 1997:04:24 -- 1997:05:22 & $\phantom{0}74.3\pm 0.6$\\
1932 & 1998:01:22 -- 1998:02:18 & $\phantom{0}87.9\pm 1.5$\\
1948 & 1999:04:04 -- 1999:05:01 & $120.4\pm 2.5$\\
1964 & 2000:06:13 -- 2000:07:10 & $188.9\pm 3.8$\\
1975 & 2001:04:09 -- 2001:05:06 & $169.9\pm 4.6$\\
1988 & 2002:03:30 -- 2002:04:26 & $196.6\pm 3.1$\\
2009 & 2003:10:18 -- 2003:11:15 & $159.9\pm 12.1$\\
\enddata
\tablenotetext{1}{ The errors are $\sigma/\sqrt{N}$, $\sigma$ being the standard deviation in the daily
10.7 cm flux during the interval covered by each data set and $N$ is the number
of days over which the daily 10.7 cm flux is averaged.}
\label{tab:tab1}
\end{deluxetable}

\clearpage

\clearpage

\clearpage

\clearpage

\clearpage

\clearpage

\clearpage

\clearpage

\clearpage

\clearpage

\clearpage

\clearpage

\clearpage

\clearpage

\clearpage

\clearpage

\clearpage

\newpage

\begin{thebibliography}{}

\bibitem[Antia \& Basu(1994)]{ab94}
Antia, H. M., \& Basu, S. 1994, A\&AS, 107, 421

\bibitem[Antia \ea(2001)]{ant01}
Antia, H. M., Basu, S., Hill, F., Howe, R., Komm, R. W., \& Schou, J. 2001,
MNRAS, 327, 1029

\bibitem[Antia \ea(2003)]{ant03}
Antia, H. M., Chitre, S. M., \& Thompson, M. J. 2003, A\&A, 399, 329

\bibitem[Bai(1990)]{ba90}
Bai, T. 1990, ApJ, 364, L17

\bibitem[Balmforth(1992)]{bal92}
Balmforth, N. J. 1992, MNRAS, 255, 632

\bibitem[Basu(2002)]{bas02}
Basu, S. 2002, in From Solar Min to Max:
Half a Solar Cycle with SOHO, Proc. SOHO 11 Symposium,
ed., A. Wilson, ESA SP-508, 7

\bibitem[Basu \& Antia(1999)]{basu-antia99}
Basu, S., \& Antia, H. M.  1999, \apj, 525, 517

\bibitem[Basu \& Antia(2000)]{basu-antia00a}
Basu, S., \& Antia, H. M.  2000, ApJ, 541, 442

\bibitem[Basu \& Antia(2002)]{ba02}
Basu, S., \& Antia, H. M.  2002, in
 From Solar Min to Max:
Half a Solar Cycle with SOHO, Proc. SOHO 11 Symposium,
ed., A. Wilson, ESA SP-508, 59

\bibitem[Basu \& Antia(2003)]{basu-antia03}
Basu, S., \& Antia, H. M.  2003, ApJ, 585, 553

\bibitem[Basu \& Mandel(2004)]{bm04}
Basu, S., \& Mandel, A. 2004, ApJ, 617, L155

\bibitem[Basu \& Thompson(1996)]{basu96}
Basu, S., \& Thompson, M. J. 1996, A\&A,  305, 631

\bibitem[Basu, Antia \& Tripathy(1999)]{basuetal99} Basu, S., Antia, H. M., \&
Tripathy, S. C. 1999, ApJ, 512, 458

\bibitem[Basu \ea(2004)]{bas04} Basu, S., Antia, H. M., \& Bogart, R. S.
2004, ApJ, 610, 1157

\bibitem[Bogart \& Basu(2004)]{bo04}
Bogart, R.S., \& Basu, S. 2004, in Proc: SOHO 14 / GONG 2004 Workshop
``Helio- and Asteroseismology: Towards a Golden Future'', ESA SP-559, 329

\bibitem[Brun \& Toomre(2002)]{br02}
Brun, A.S., Toomre, J. 2002, ApJ, 570, 865

\bibitem[Brun et al.(2004)]{br04}
Brun, A.S., Miesch, M., Toomre, J. 2004, ApJ, 614, 1073

\bibitem[Chou \& Serebryanskiy(2005)]{chou05}
Chou, D.-Y., \& Serebryanskiy, A. 2005, ApJ, 624, 420

\bibitem[Cox \& Kidman]{cox84}
Cox, A. N., \& Kidman, R. B. 1984,
in  Theoretical problems in stellar stability and oscillations
(Li\`ege: Institut d'Astrophysique), 259

\bibitem[Dappen et al.(1991)]{dap91}
D\"appen, W., Gough, D. O., Kosovichev, A. G., \& Thompson, M. J.
1991, in Challenges to Theories of the Structure of Moderate-Mass
Stars, eds., D. Gough \& T. Toomre,
Lecture notes in Physics, 388, 111

\bibitem[de Toma et al.(2000)]{detoma00}
de Toma, G., White, O. R., \& Harvey, K. L. 2000, ApJ, 529, 1101

\bibitem[Duchlev & Dermendjiev(1996)]{du96}
Duchlev, P. I., \& Dermendjiev, V. N. 1996, Sol. Phys., 168, 205 

\bibitem[Duvall et al.(1993)]{duvetal93}
Duvall, T. L., Jr., Jefferies, S. M., Harvey, J. W., \& Pomerantz, M. A.
1993, Nature, 362, 430

\bibitem[Dziembowski et al.(1990)]{dz90}
Dziembowski, W. A., Pamyatnykh, A. A., \& Sienkiewicz, R. 1990,
MNRAS, {244}, 542

\bibitem[Dziembowski et al.(1994)]{dz94}
Dziembowski, W. A., Goode, P. R., Pamyatnykh, A. A., \& Sienkiewicz, R. 1994,
\apj, {432}, 417

\bibitem[Eff-Darwich \ea(2002)]{eff02}
Eff-Darwich, A., Korzennik, S. G., Jim\'enez-Reyes, S. J., \&
P\'erez Hern\'andez, F. 2002, ApJ, 580, 574


\bibitem[Gough \ea(1996)]{dog96}
Gough, D. O., Kosovichev, A. G., Toomre, J. et al., 1996, Science, 272, 1296

\bibitem[Haber et al.(2002)]{ha02}
Haber, D. A., Hindman, B. W., Toomre, J., Bogart, R. S., Larsen. R. M., \& Hill, F. 2002,
ApJ, 570, 855

\bibitem[Hill(1988)]{hill88}
Hill, F. 1988, \apj, 333, 996

\bibitem[Hindman et al.(2000)]{hind00}
Hindman, B. W., Haber, D., Toomre, J., \& Bogart, R. 2000, Sol.\ Phys., 192, 363

\bibitem[Howard(1974)]{ho74}
Howard, R. 1974, Sol. Phys., 38, 59

\bibitem[Howe et al.(2000)]{ho00}
Howe R., Christensen-Dalsgaard J., Hill F., Komm R. W.,
Larsen R. M., Schou J., Thompson M. J., \& Toomre J. 2000,  ApJ
533, L163

\bibitem[Howe  et al.(2004a)]{ho04a}
Howe, R., Komm, R. W., Hill, F., Christensen-Dalsgaard, J., Haber, D. A., Schou, J., 
\& Thompson, M. J. 2004a, in Proc. SOHO 14/ GONG 2004 11 ``Helio- and Asteroseismology: Towards 
a Golden Future'' ESA SP-559, 472

\bibitem[Howe  et al.(2004b)]{ho04b}
Howe, R., Komm, R. W., Hill, F., Haber, D. A., \& Hindman, B. W. 2004b, ApJ 608, 562

\bibitem[Knaack et al.(2004)]{kn04}
Knaack, R., Stenflo, J. O., \& Berdyugina, S. V. 2004, A\&A, 418, L17 

\bibitem[Korzennik \ea(2004)]{ko04}
Korzennik, S. G., Rabello-Soares, M. C., \& Schou, J. 2004, ApJ, 602, 481

\bibitem[Kosovichev \ea(2000)]{kos00}
Kosovichev, A. G., Duvall, T. L., Jr., \& Scherrer, P. H. 2000,
Solar Phys., 192, 159

\bibitem[Kosovichev \ea(2001)]{kos01}
Kosovichev, A. G., Duvall, T. L., Jr., Birch, A. C., Gizon, L.,
Scherrer, P. H., \& Zhao, J. 2001, in Helio- and Asteroseismology at
the Dawn of the Millennium, Proc. SOHO 10/GONG 2000 Workshop,
ed., A. Wilson, ESA SP-464, 701

\bibitem[Miesch et al.(2006)]{mi06}
Miesch, M.S., Brun, A.S., Toomre, J. 2006, ApJ, 641, 618

\bibitem[Patr\'on et al.(1997)]{patron-etal97} Patr\'on, J., {\it et al.} 1997,
ApJ, 485, 869

\bibitem[Pijpers \& Thompson(1992)]{pij92}
Pijpers, F. P., \& Thompson, M. J. 1992, A\&A,  262, L33

\bibitem[Pijpers \& Thompson(1994)]{pij94}
Pijpers, F. P., \& Thompson, M. J. 1994, A\&A,  281, 231

\bibitem[Rabello-Soares \ea(1999)]{rab99}
Rabello-Soares, M. C., Basu, S., \& Christensen-Dalsgaard, J. 1999,
MNRAS,  309, 35

\bibitem[Rajaguru et al.(2001)]{raj01}
Rajaguru, S. P., Basu, S., \& Antia, H. M. 2001, ApJ, 563, 410

\bibitem[Schou(1999)]{sch99}
Schou, J., 1999, ApJ, 523, L181

\bibitem[Schou \ea(1998)]{sch98}
Schou, J., Antia, H. M., Basu, S. et al., 1998, ApJ, 505, 390

\bibitem[Sekii(1997)]{se97}
Sekii, T. 1997, in Proc. IAU Symp. 181: Sounding Solar and Stellar Interiors,
eds. J Provost, F.-X. Schmider, Dordrecht: Kluwer, 189

\bibitem[Thompson \ea(1996)]{mjt96}
Thompson, M. J., Toomre, J., Anderson, E. R. et al., 1996, Science, 272, 1300

\bibitem[Thompson \ea(2003)]{mjt03}
Thompson, M. J., Christensen-Dalsgaard, J., Miesch, M.S., Toomre, J. 2003,
ARA\&A, 41, 599

\bibitem[Verner \ea(2006)]{v06}
Verner, G.A., Chaplin, W.J., Elsworth, Y. 2006, ApJ, 640, L95


\end{thebibliography}
\end{document}